\documentclass[3p,times,10pt]{elsarticle}
\usepackage[compress]{numcompress}
\usepackage{amssymb}
\usepackage{amsmath}
\usepackage{graphicx}
\usepackage{epsfig}
\usepackage{ctable}

\begin{document}

\begin{frontmatter}

\title{Scaling behavior and phase diagram of a $\mathcal{PT}$-symmetric
non-Hermitian Bose-Hubbard system}
\author{L.~Jin}
\author{Z.~Song\corref{cor1}}
\ead{songtc@nankai.edu.cn}
\cortext[cor1]{Corresponding author}
\address{School of Physics, Nankai University - Tianjin 300071, China}

\begin{abstract}
We study scaling behavior and phase diagram of a $\mathcal{PT}$%
-symmetric non-Hermitian Bose-Hubbard model. In the free interaction
case, using both analytical and numerical approaches, the metric operator
for many-particle is constructed. The derived properties of the metric
operator, similarity matrix and equivalent Hamiltonian reflect the fact that
all the matrix elements change dramatically with diverging derivatives near
the exceptional point. In the nonzero interaction case, it is found
that even small on-site interaction can break the $\mathcal{PT}$ symmetry
drastically. It is demonstrated that the scaling law can be established for
the exceptional point in both small and large interaction limit. Based on
perturbation and numerical methods, we also find that the phase diagram shows rich structure: there exist multiple regions of
unbroken $\mathcal{PT}$ symmetry.
\end{abstract}


\end{frontmatter}



\section{Introduction}

\label{sec_introduction}

Non-Hermitian Hamiltonian is traditionally used to describe open system
phenomenologically. It has profound applications in nuclear physics, quantum
transport, quantum chemistry, as well as in quantum optics~\cite%
{nonHermitianH}. Since the discovery of a parity-time ($\mathcal{PT}$)
symmetric non-Hermitian Hamiltonian can still have an entirely real
spectrum, extensive efforts were paid to the pseudo-Hermitian quantum theory~%
\cite%
{Geyer1992,AliM,AM37,AM38,Ahmed,Berry,EPsHeiss,Dembowski,Heiss40,Bender98,Bender,BenderRPP,MZnojil,Jones,Tateo,Mueller,Longhi,LonghiPRL103,PTinOptics,PTinOpticsPO,Klaiman,AGuo,JLPT,Ghosh}%
, which paved the way to our understanding of the connection between
non-Hermitian systems and the real physical world. In general, a $\mathcal{PT%
}$-symmetric non-Hermitian Hamiltonian has unbroken as well as broken $%
\mathcal{PT}$-symmetric phases, the phase boundary is referred to as the
exceptional points (EPs). Studies of the EPs were presented theoretically
and experimentally over a decade ago~\cite{Berry,EPsHeiss,Dembowski}.
Recently, the experimental realization of $\mathcal{PT}$-symmetric systems
in optics were suggested through creating a medium with alternating regions
of gain and loss~\cite{PTinOptics,PTinOpticsPO,Klaiman}, in which the
complex refractive index satisfies the condition $V\left( x\right) =V^{\ast
}\left( -x\right) $ and $\mathcal{PT}$ symmetry breaking was observed~\cite%
{AGuo}.

One of the characteristic features of the $\mathcal{PT}$-symmetric system is
the ubiquitous phase diagram which depicts the symmetry of the
eigenfunctions and the reality of the spectrum~\cite{Bender98}. The phase
separation arises from the fact that although $H$ and the $\mathcal{PT}$
operator commute, the eigenstates of $H$ may or may not be eigenstates of
the $\mathcal{PT}$ operator, since the $\mathcal{PT}$ operator is not
linear. In the broken $\mathcal{PT}$-symmetric phase the spectrum becomes
partially or completely complex, while in the unbroken $\mathcal{PT}$%
-symmetric phase both $H$ and $\mathcal{PT}$ share the same set of
eigenvectors and the spectrum is entirely real. Recently, the phase diagram
of a lattice model has been investigated. It is shown that the critical
point is sensitive to the distribution of the coupling constant and on-site
potential~\cite{Giorgi,Bendix,Joglekar,JoglekarDegree}.

In this paper, we investigate the effect of on-site interaction on the phase
boundary of a $\mathcal{PT}$-symmetric Bose-Hubbard system. Our approach is
based on our previous work in Ref.~\cite{JLPT}, where we have systematically
investigated an $N$-site tight-binding chain with a pair of conjugate
imaginary potentials $\pm i\gamma $ located at edges. Here we will
generalize this description by considering many-particle system and adding
the on-site Hubbard interaction $U$. In the free interaction case,
many-particle eigenstates are obtained in aid of the single-particle
solutions. We also construct the metric operator to investigate the
Hermitian counterpart and observables in the framework of complex quantum
mechanics. In nonzero $U$ case, we restrict our attention to the influence
of the nonlinear on-site interaction $U$ on the boundary between unbroken
and broken $\mathcal{PT}$-symmetric phases. Exact Bethe ansatz solution and
numerical results show that small on-site interaction can reduce the
critical point $\gamma _{c}$ drastically. Moreover, numerical results show
that there exist multiple regions of unbroken $\mathcal{PT}$ symmetry and
the number of such regions increases as the system size $N$ increases.

This paper is organized as follows. Section \ref{sec_PT_symmetric_BH_model}
describes the Hamiltonian of a $\mathcal{PT}$-symmetric Bose-Hubbard model.
In Section \ref{sec_interaction_free}, we focus on the interaction-free
case. Based on the single-particle solutions, we construct the many-particle
eigenstates and metric operator to study the Hermitian counterpart and
observables. Section \ref{sec_nonzero_interaction} is devoted to the case of
nonzero interaction. Based on the approximation solutions, we investigate
the critical scaling behavior and the phase diagram. Our findings are
briefly summarized and the physical relevance of the model and results are
discussed\ in Section \ref{sec_conclusion}.

\section{$\mathcal{PT}$-Symmetric Bose-Hubbard model}

\label{sec_PT_symmetric_BH_model} The Bose-Hubbard model\ gives an
approximate description of the physics of interacting bosons on a lattice.
Since it embodies essential features of ultracold atoms in optical lattices,
the Bose-Hubbard model plays an important role in quantum many-body physics~%
\cite{Greiner,Jaksch}. Optical realization of two-site Bose-Hubbard model in
coupled cavity arrays and waveguides have been proposed~\cite%
{Hartmann,QBEC,LonghiJPB44}. It is worth noting that non-Hermitian
Bose-Hubbard dimer has attracted enormous research attention in recent years~%
\cite{Hiller,GraefeJPA41,Graefe08PRL,GraefePRA82,HXiong,HZhong}. Theoretical
investigations on two site open Bose-Hubbard system was firstly presented in~%
\cite{Hiller}. For a $\mathcal{PT}$-symmetric non-Hermitian Bose-Hubbard
dimer, the spectrum and the exceptional points were studied in~\cite%
{GraefeJPA41}. After that, dynamics in a leaking double well trap described
by non-Hermitian Bose-Hubbard Hamiltonian with additional decay term was
investigated under the mean field approximation~\cite{Graefe08PRL}. Through
dynamical study of Bose-Einstein condensed gases, it was shown that
imaginary periodic potential may induce perfect quantum coherence between
two different condensates~\cite{HXiong}. The realization of such open system
can be put into practice as a BEC in a double well trap, where the
condensate could escape from the traps via tunneling. Most investigations
mainly focus on the Bose-Hubbard model with effective decay term in one
site.However, it should be noticed that non-Hermitian Bose-Hubbard dimer
with complex coupling terms has also attracted some attention, the decay of
quantum states could be controlled by modulating the particle-particle
interaction strength and the dissipation in the tunneling process~\cite%
{HZhong}. On the other hand, the Bose-Hubbard model with particle loss was
investigated in an alternative way through employing Lindblad\ master
equation \cite{Kepesidis}, which phenomenologically describes non-unitary
evolution of an open system \cite{Lindblad}. Recently, $\mathcal{PT}$%
-symmetric quantum Liouvillean dynamics is also investigated \cite{Prosen}.
In this paper, we investigate the property of a\ non-Hermitian Hamiltonian
in the framework of quantum mechanics.

Nevertheless, although there have been no experiments to show clearly and
definitively that a finite non-Hermitian Hamiltonian do exist in nature,
many interesting features have been observed in non-Hermitian optical
systems, such as double refraction, power oscillations, nonreciprocal
phenomenon, etc.~\cite{LonghiPRL103,PTinOptics,PTinOpticsPO,Klaiman}. So
far, most contributions to pseudo-Hermitian quantum theory were for the
single particle problem. Particularly, a two-mode $\mathcal{PT}$-symmetric
non-Hermitian Bose-Hubbard system with an imaginary potential on the edge
has been investigated~\cite{GraefeJPA41}. In this paper, we focus on the
influence of on-site Hubbard interaction, not restricted to a dimer but to
two-particle problem of an $N$-site Bose-Hubbard system. We mainly study the
$\mathcal{PT}$-symmetric non-Hermitian Bose-Hubbard system. The Hamiltonian
reads
\begin{equation}
H=-J\sum_{l=1}^{N-1}\left( a_{l}^{\dag }a_{l+1}+\text{\textrm{H.c.}}\right) +%
\frac{U}{2}\sum_{l=1}^{N}a_{l}^{\dag 2}a_{l}^{2}+i\gamma \left(
n_{1}-n_{N}\right)  \label{H_U}
\end{equation}%
where $a_{l}^{\dag }$ is the creation operator of the boson at the $l$th
site and the tunneling strength is denoted by $J$. The on-site interaction
strength and the on-site potential are denoted by $U$ and $i\gamma $,
respectively. $H$ is a $\mathcal{PT}$-symmetric Hamiltonian, i.e., $[%
\mathcal{PT},H]=0$, where the action of the parity operator $\mathcal{P}$ is
defined by $\mathcal{P}:l\rightarrow N+1-l$ and the time-reversal operator $%
\mathcal{T}$ by $\mathcal{T}:i\rightarrow -i$. Both single-particle solution
and the critical point $\gamma _{c}$ for interaction-free Hamiltonian $H^{%
\text{\textrm{free}}}=H(U=0)$\ have been obtained explicitly in our previous
study~\cite{JLPT}. The main goal of the present work is to study the
influence of the nonlinear interaction on the features of the system.

\section{Interaction-free system}

\label{sec_interaction_free}When dealing with a $\mathcal{PT}$-symmetric
non-Hermitian system, to our knowledge, most researchers concerned about the
single-particle problem, since it is believed that the extension of this
study to a many-body problem is straightforward. Nevertheless, due to the
particular formalism of non-Hermitian quantum mechanics, it is worthwhile to
investigate the many-particle system with $U=0$. In the following, we will
extend the obtained results for single particle to the case of many-particle
system with zero $U$.

\subsection{Many-particle solutions}

In a non-Hermitian system, although the particle probability is no long
conservative, the particle number
\begin{equation}
\hat{N}_{p}=\sum_{l=1}^{N}a_{l}^{\dag }a_{l}  \label{N_sum}
\end{equation}%
still shares the common eigenfunctions with the Hamiltonian due to the
commutation relation
\begin{equation}
\lbrack \hat{N}_{p},H^{\text{\textrm{free}}}]=[\hat{N}_{p},H]=0.
\label{N comm}
\end{equation}%
This fact indicates that the proper inner product should accord with the
conservation of particle number. Therefore the eigenstates of $H^{\text{%
\textrm{free}}}$\ or $H$\ can be obtained in each invariant subspace $%
V^{N_{p}}$, which is spanned by the occupation number basis%
\begin{equation}
\left\vert n_{1},n_{2},...,n_{N}\right\rangle \equiv
\prod_{i=1}^{N}\left\vert n_{i}\right\rangle ,  \label{basis_coordinate}
\end{equation}%
with $\hat{N}_{p}=\sum_{l=1}^{N}\hat{n}_{l}$, where $\left\vert
n_{l}\right\rangle \equiv (a_{l}^{\dag })^{n_{l}}/\sqrt{n_{l}!}\left\vert
\text{vac}\right\rangle $. Notice that $\left\{ \left\vert
n_{1},n_{2},...,n_{N}\right\rangle \right\} $ is orthonormal set under the
Dirac inner product.\ According to our previous work~\cite{JLPT}, the
single-particle solutions $\{\left\vert \phi _{+}^{k}\right\rangle \}$ and $%
\{\left\vert \phi _{-}^{k}\right\rangle \}$, which are eigenfunctions of the
systems $H^{\text{\textrm{free}}}$ and $(H^{\text{\textrm{free}}})^{\dag }$%
,\ i.e. $\left\vert \phi _{+}^{k}\right\rangle
=\sum_{l}f_{k}^{l}a_{l}^{\dagger }\left\vert \text{vac}\right\rangle $, $%
\left\vert \phi _{-}^{k}\right\rangle =\sum_{l}g_{k}^{l}a_{l}^{\dagger
}\left\vert \text{vac}\right\rangle $. They can construct the biorthogonal
basis set, i.e.,%
\begin{equation}
\begin{aligned} \sum_{l}f_{k}^{l}(g_{k^{\prime }}^{l})^{\ast } &=\delta
_{kk^{\prime }}, \\ \sum_{k}f_{k}^{l}(g_{k}^{l^{\prime }})^{\ast } &=\delta
_{ll^{\prime }}, \end{aligned}  \label{biorthogonal}
\end{equation}%
here $f_{k}^{l}=\phi _{+}^{k}\left( l\right) $, $g_{k}^{l}=\phi
_{-}^{k}\left( l\right) $ have the form%
\begin{equation}
\phi _{\pm }^{k}\left( l\right) =\frac{e^{ik\left( l-N_{0}\right) }-\xi
_{\pm }e^{-ik\left( l+N_{0}\right) }}{\left\vert \sqrt{\left[ 1+\left\vert
\xi _{\pm }\right\vert ^{2}\right] \sin \left( Nk\right) /\sin k-2N\xi _{\pm
}e^{-ik\left( N+1\right) }}\right\vert },  \label{phi+/-}
\end{equation}%
where%
\begin{equation}
\xi _{\pm }\left( k\right) =\frac{\gamma e^{ik}\mp iJ}{\gamma e^{-ik}\mp iJ}
\label{xi}
\end{equation}%
and $N_{0}=\left( N+1\right) /2$. The symmetries of the wavefunctions and
the spectrum reveal that\ there are two phases, unbroken and broken phase,
which are separated by the critical point $\gamma _{c}$,
\begin{equation}
\gamma _{c}=\left\{
\begin{array}{l}
\pm J\text{, }N=2n \\
\pm J\sqrt{\frac{n+1}{n}}\text{, }N=2n+1%
\end{array}%
\right. ,
\end{equation}%
where $n=1,2,...$ In the unbroken region with $\left\vert \gamma \right\vert
<\left\vert \gamma _{c}\right\vert $,\ all the solutions possess $\mathcal{PT%
}$ symmetry
\begin{equation}
\mathcal{PT}f_{k}^{l}=\left( f_{k}^{N+1-l}\right) ^{\ast }=f_{k}^{l},
\label{PTf_k}
\end{equation}%
and the spectrum is entirely real. In the following, we will demonstrate
that the above analysis can be extended to many-particle sector. Actually,
one can define the operators in $k$ space in the form of%
\begin{equation}
\begin{aligned} \bar{a}_{k} &=\sum_{l}f_{k}^{l}a_{l}^{\dagger }, \\ a_{k}
&=\sum_{l}\left( g_{k}^{l}\right) ^{\ast }a_{l}, \end{aligned}  \label{a_k}
\end{equation}%
which obey the standard bosonic commutation relations
\begin{equation}
\begin{aligned} \left[ a_{k},\bar{a}_{k^{\prime }}\right] &=\delta
_{kk^{\prime }},\\ \left[ a_{k},a_{k^{\prime }}\right] &=\left[
\bar{a}_{k},\bar{a}_{k^{\prime}}\right] =0. \end{aligned}  \label{cannonical}
\end{equation}%
Then the Hamiltonian $H^{\text{\textrm{free}}}$\ can be written as the
diagonal form
\begin{equation}
H^{\text{\textrm{free}}}=\sum_{k}\epsilon _{k}\bar{a}_{k}a_{k},
\label{H_free}
\end{equation}%
where $\epsilon _{k}=-2J\cos k$ is real. With respect to the canonical
commutation relations of (\ref{cannonical}), the Hamiltonian in the form of (%
\ref{phi+/-}) can be regarded as the term of the so-called second
quantization representation. Defining the occupation number state in $k$%
-space as%
\begin{equation}
\begin{aligned} \overline{\left\vert n_{k_{i}}\right\rangle }\equiv
\frac{\left( \bar{a}_{k_{i}}\right)
^{n_{k_{i}}}}{\sqrt{n_{k_{i}}!}}\left\vert \text{vac}\right\rangle, \\
\left\vert n_{k_{i}}\right\rangle \equiv \frac{(a_{k_{i}}^{\dag
})^{n_{k_{i}}}}{\sqrt{n_{k_{i}}!}}\left\vert \text{vac}\right\rangle,
\end{aligned}  \label{n_k}
\end{equation}%
which satisfy%
\begin{equation}
\begin{aligned} \bar{a}_{k_{i}}\overline{\left\vert n_{k_{i}}\right\rangle }
&=\sqrt{n_{k_{i}}+1}\overline{\left\vert n_{k_{i}}+1\right\rangle}, \\
a_{k_{i}}\overline{\left\vert n_{k_{i}}\right\rangle }
&=\sqrt{n_{k_{i}}}\overline{\left\vert n_{k_{i}}-1\right\rangle }, \\
a_{k_{i}}^{\dag }\left\vert n_{k_{i}}\right\rangle
&=\sqrt{n_{k_{i}}+1}\left\vert n_{k_{i}}+1\right\rangle , \\
\bar{a}_{k_{i}}^{\dag }\left\vert n_{k_{i}}\right\rangle
&=\sqrt{n_{k_{i}}}\left\vert n_{k_{i}}-1\right\rangle . \end{aligned}
\label{four_a_operators}
\end{equation}%
Then, the eigenstates in the subspace $V^{N_{p}}$ of $H^{\text{\textrm{free}}%
}$\ and $(H^{\text{\textrm{free}}})^{\dag }$\ read%
\begin{equation}
\begin{aligned} \overline{\left\vert
n_{k_{1}},n_{k_{2}},...,n_{k_{N}}\right\rangle } &\equiv
\prod\limits_{i=1}^{N}\overline{\left\vert n_{k_{i}}\right\rangle },\\
\left\vert n_{k_{1}},n_{k_{2}},...,n_{k_{N}}\right\rangle &\equiv
\prod\limits_{i=1}^{N}\left\vert n_{k_{i}}\right\rangle , \end{aligned}
\label{Eigenstate}
\end{equation}%
respectively. They correspond to the same eigenvale as%
\begin{equation}
E\left( n_{k_{1}},n_{k_{2}},...,n_{k_{N}}\right)
=\sum_{l=1}^{N}n_{k_{l}}\epsilon _{k_{l}}.  \label{E}
\end{equation}%
and the total particle number as $N_{p}=\sum_{l=1}^{N}n_{k_{i}}$.
Equivalently, we have%
\begin{eqnarray}
H^{\text{\textrm{free}}}\overline{\left\vert
n_{k_{1}},n_{k_{2}},...,n_{k_{N}}\right\rangle } &=&E\left(
n_{k_{1}},n_{k_{2}},...,n_{k_{N}}\right) \overline{\left\vert
n_{k_{1}},n_{k_{2}},...,n_{k_{N}}\right\rangle },  \label{H_free_occupa} \\
\left( H^{\text{\textrm{free}}}\right) ^{\dag }\left\vert
n_{k_{1}},n_{k_{2}},...,n_{k_{N}}\right\rangle &=&E\left(
n_{k_{1}},n_{k_{2}},...,n_{k_{N}}\right) \left\vert
n_{k_{1}},n_{k_{2}},...,n_{k_{N}}\right\rangle .  \label{H_free_dagger}
\end{eqnarray}%
Thus we conclude that for many-particle case, the phase boundary is still at
$\gamma _{c}$. Notice that the eigenstates $\{\overline{\left\vert
n_{k_{1}},n_{k_{2}},...,n_{k_{N}}\right\rangle }\}$, $\left\{ \left\vert
n_{k_{1}},n_{k_{2}},...,n_{k_{N}}\right\rangle \right\} $\ construct a
biorthogonal set instead of the set $\left\{ \left\vert
n_{1},n_{2},...,n_{N}\right\rangle \right\} $ under the Dirac inner product.

\subsection{Metric and Hermitian counterpart}

\begin{table}[tbh]
\caption{The matrix representation of the metric operator $\protect\eta $,
similarity matrix $\protect\rho $, and the equivalent Hermitian Hamiltonian $%
h$ for systems with $N=2,$ $3,$ and $4$ are listed.\ Here we denote $\protect%
\lambda =\protect\sqrt{J^{2}-\protect\gamma ^{2}},$ $\protect\varsigma _{\pm
}=\protect\sqrt{J\pm \protect\gamma },$ and $\protect\tau =\protect\sqrt{%
J^{2}-\protect\gamma ^{2}/2}$. These matrices satisfy the relations (\protect
\ref{MOs_rela}). The analytical expression for the cases of $N=2,3$ and the
numerical plot in Fig. \protect\ref{fig_MOs} for the case of $N=4$ show that
the derivatives of all the matrix elements with respect to $\protect\gamma $%
\ diverge at the exceptional points.}
\label{table}\centering
\setlength{\tabcolsep}{0.3em} 
\begin{tabular}{cccc}
\toprule $N$ & $2$ & $3$ & $4$ \\ \hline
$\gamma _{c}$ & $1$ & $\sqrt{2}$ & $1$ \\
$H$ & $\left(
\begin{array}{cc}
i\gamma & -J \\
-J & -i\gamma%
\end{array}%
\right) $ & $\left(
\begin{array}{ccc}
i\gamma & -J & 0 \\
-J & 0 & -J \\
0 & -J & -i\gamma%
\end{array}%
\right) $ & $\left(
\begin{array}{cccc}
i\gamma & -J & 0 & 0 \\
-J & 0 & -J & 0 \\
0 & -J & 0 & -J \\
0 & 0 & -J & -i\gamma%
\end{array}%
\right) $ \\
$\eta $ & $\frac{1}{\lambda }\left(
\begin{array}{cc}
J & i\gamma \\
-i\gamma & J%
\end{array}%
\right) $ & $\frac{1}{\tau ^{2}}\left(
\begin{array}{ccc}
J^{2} & i\gamma J & -\gamma ^{2}/2 \\
-i\gamma J & J^{2}+\gamma ^{2}/2 & i\gamma J \\
-\gamma ^{2}/2 & -i\gamma J & J^{2}%
\end{array}%
\right) $ & $\left(
\begin{array}{cccc}
\alpha & -i\beta & \mu & -i\nu \\
i\beta & \chi & -i\beta & \mu \\
\mu & i\beta & \chi & -i\beta \\
i\nu & \mu & i\beta & a%
\end{array}%
\right) $ \\
$\rho $ & $\frac{1}{2\sqrt{\lambda }}\left(
\begin{array}{cc}
\varsigma _{+}+\varsigma _{-} & i\varsigma _{+}-i\varsigma _{-} \\
i\varsigma _{-}-i\varsigma _{+} & \varsigma _{+}+\varsigma _{-}%
\end{array}%
\right) $ & $\frac{1}{2\tau }\left(
\begin{array}{ccc}
\tau +J & i\gamma & \tau -J \\
-i\gamma & 2J & i\gamma \\
\tau -J & -i\gamma & \tau +J%
\end{array}%
\right) $ & $\left(
\begin{array}{cccc}
a & -ib & c & -id \\
ib & r & -is & c \\
c & is & r & -ib \\
id & c & ib & a%
\end{array}%
\right) $ \\
$h$ & $\left(
\begin{array}{cc}
0 & -\lambda \\
-\lambda & 0%
\end{array}%
\right) $ & $\left(
\begin{array}{ccc}
0 & -\tau & 0 \\
-\tau & 0 & -\tau \\
0 & -\tau & 0%
\end{array}%
\right) $ & $\left(
\begin{array}{cccc}
0 & x & 0 & y \\
x & 0 & z & 0 \\
0 & z & 0 & x \\
y & 0 & x & 0%
\end{array}%
\right) $ \\
\bottomrule &  &  &
\end{tabular}
\end{table}
According to the quasi-Hermitian quantum mechanics, a bounded
positive-definite Hermitian operator $\eta $\ in each invariant subspace can
be constructed~\cite{AM37} via the eigenstates of $(H^{\text{\textrm{free}}%
})^{\dag }$ as
\begin{equation}
\eta =\sum_{\{n_{k_{i}}\}}\left\vert
n_{k_{1}},n_{k_{2}},...,n_{k_{N}}\right\rangle \left\langle
n_{k_{1}},n_{k_{2}},...,n_{k_{N}}\right\vert ,  \label{metric}
\end{equation}%
which is called the metric operator to define the biorthogonal inner
product. The $\eta $-metric operator inner product leads to a unitary time
evolution \cite{AM37,Bender98}. Here $\{n_{k_{i}}\}$\ denotes all the
possible states with $\sum_{k_{i}}n_{k_{i}}=N_{p}$. One can see that the
metric operator fulfils
\begin{equation}
\eta H^{\text{\textrm{free}}}\eta ^{-1}=(H^{\text{\textrm{free}}})^{\dag },
\label{pseudo}
\end{equation}%
and thus can be employed to\ construct a Hermitian Hamiltonian $h$\ that
possesses the same spectrum as $H^{\text{\textrm{free}}}$. Actually, the
matrix representation of $\eta $ based on the orthonormal basis under the
Dirac inner product, says (\ref{basis_coordinate}),\ shows that it is a
Hermitian matrix. Furthermore, let $\rho =\sqrt{\eta }$ be the unique
positive-definite square root of $\eta $. Then the Hermitian operator $\rho $
acts as a similarity transformation to map the non-Hermitian Hamiltonian $H^{%
\text{\textrm{free}}}$\ onto its equivalent Hermitian counterpart $h$ by%
\begin{equation}
h=\rho H\rho ^{-1}.  \label{h}
\end{equation}

To demonstrate such a procedure we take the small size systems as examples.
In the following, we consider the Hamiltonian matrices $H_{N}$\ in
single-particle subspace for chain systems with $N=2$, $3$, and $4$. We
derive the explicit forms of metric operator $\eta $, similarity matrix $%
\rho $, and Hermitian counterpart $h$ for non-Hermitian Hamiltonian $H_{N}$.
The matrices $\eta $, $\rho $, and $h$ for systems $N=2$, $3$ are expressed
in analytical forms in Table \ref{table}, while the ones for $N=4$ are
plotted in Fig. \ref{fig_MOs}. It is noticed that they satisfy the following
relations
\begin{subequations}
\label{MOs_rela}
\begin{eqnarray}
\mathcal{R}\eta \mathcal{R} &=&\eta ^{\ast }=\eta ^{-1},  \label{MOs_a} \\
\mathcal{R}\rho \mathcal{R} &=&\rho ^{\ast }=\rho ^{-1},  \label{MOs_b} \\
\mathcal{PT}\eta \mathcal{PT} &=&\eta ,\mathcal{PT}\rho \mathcal{PT}=\rho ,
\label{MOs_c}
\end{eqnarray}%
where matrix $\mathcal{R}$ is defined as $\mathcal{R}\left( m,n\right)
=\left( -1\right) ^{m}\delta _{mn}$.
\begin{figure}[tbh]
\includegraphics[bb=27 190 554 609, width=5.5 cm, clip]{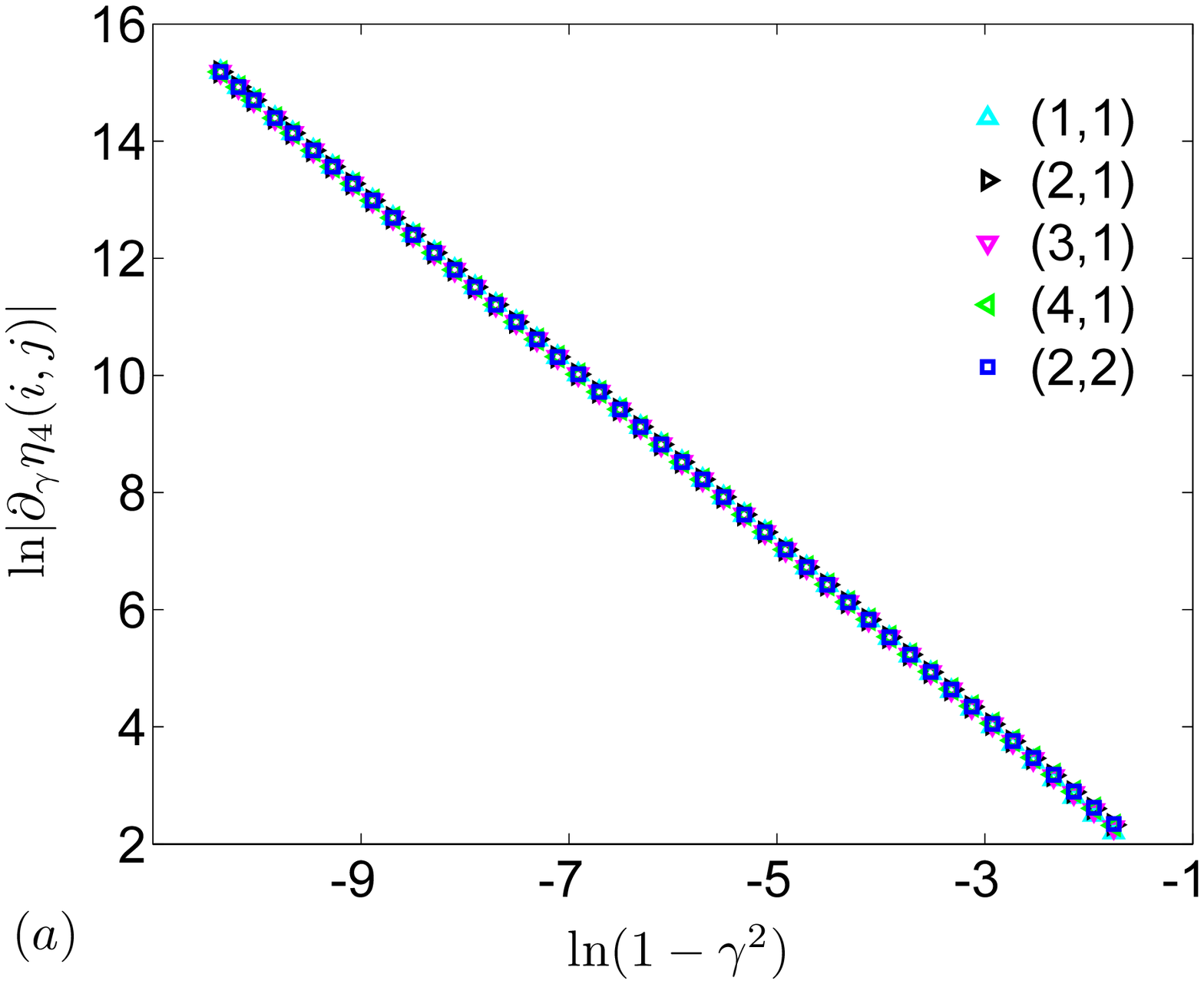} %
\includegraphics[bb=27 190 554 609, width=5.5 cm, clip]{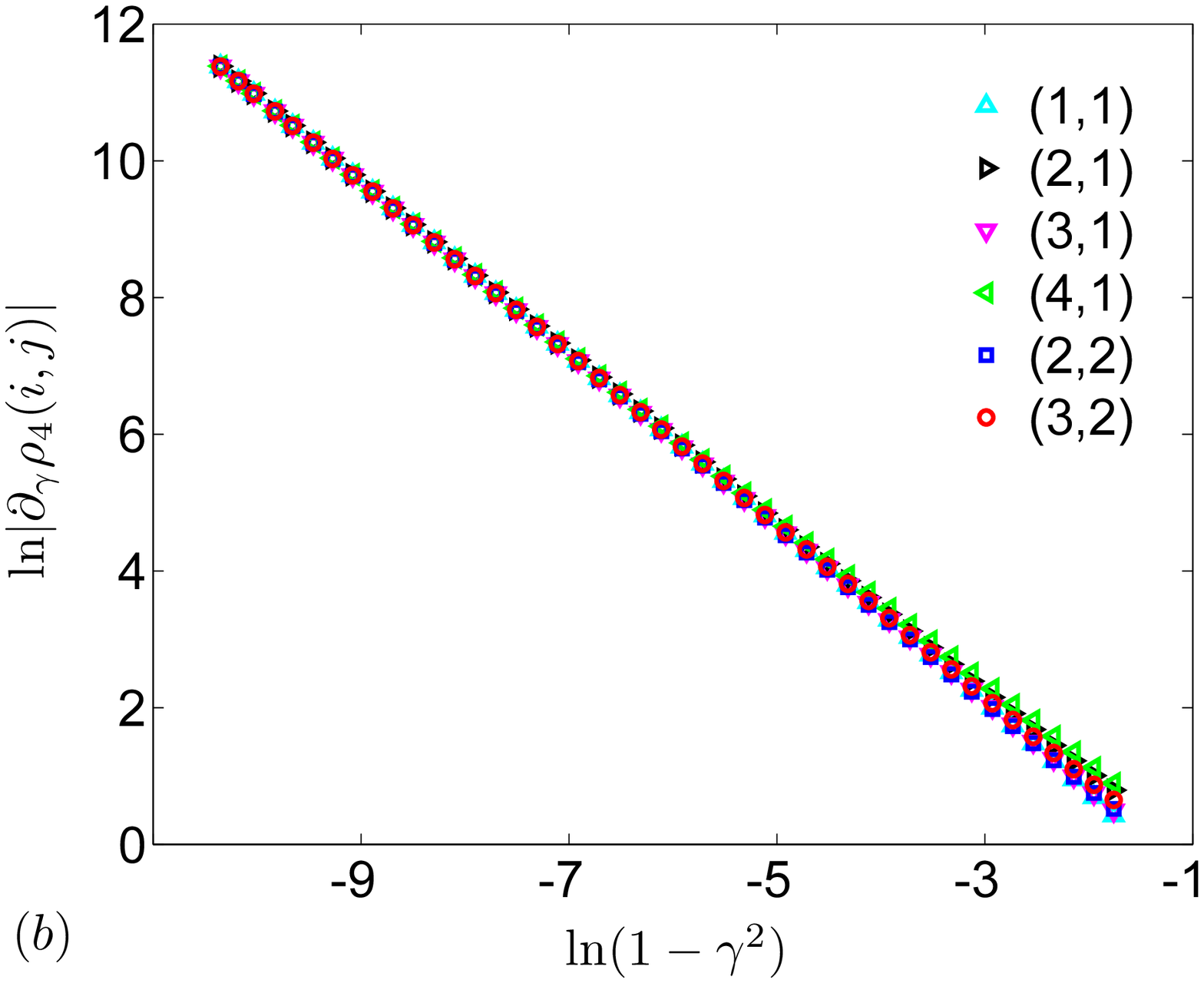} %
\includegraphics[bb=27 190 554 609, width=5.5 cm, clip]{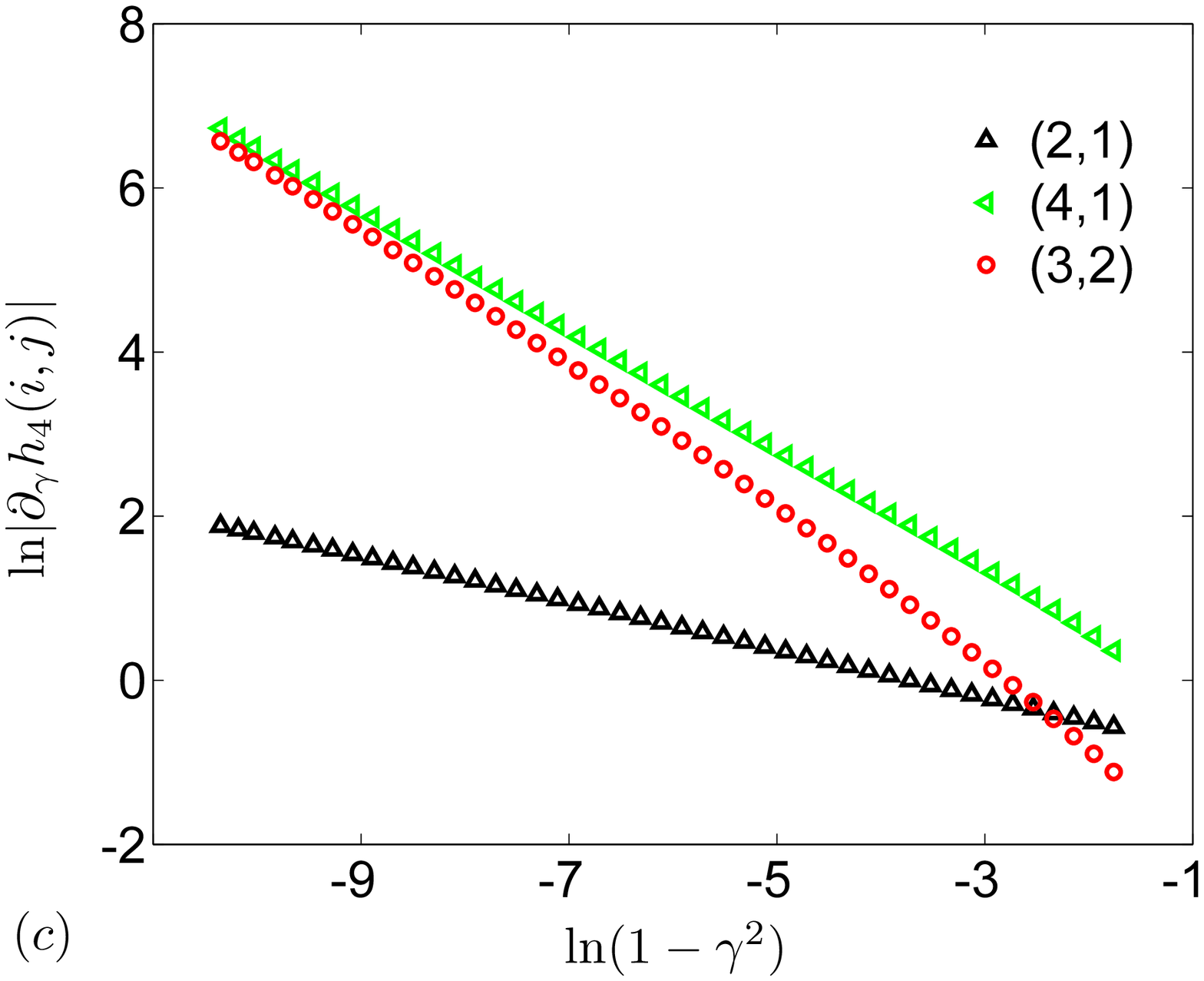}
\caption{Plots of the derivatives with respect to $\protect\gamma $ of
matrix elements\ of $\protect\eta $ (left panel), $\protect\rho $ (middle
panel), and $h$\ (right panel) for the case of $N=4$\ near the exceptional
point $\protect\gamma \rightarrow \protect\gamma _{c}=1$\ (taking $J=1$).
The numerical results are obtained by exact diagonalization, where $(i,j)$\
denotes the index of the matrix elements. It shows that the derivatives of
the elements diverge as $\protect\gamma \rightarrow \protect\gamma _{c}$.}
\label{fig_MOs}
\end{figure}

All the matrices have the common features: the derivatives of them with
respect to $\gamma $\ diverge at the exceptional points. This result is not
surprising since there is at least a pair of energy levels exhibit repulsion
characteristic near $\gamma _{c}$. Nevertheless, we notice that the
derivative of the original non-Hermitian Hamiltonian $H^{\text{\textrm{free}}%
}$\ is always finite, which reveals the essential difference between a
pseudo-Hermitian Hamiltonian and its equivalent Hermitian counterpart. The
physics of $h$\ near the exceptional point is also obvious: the coupling
constants of Hermitian counterpart\ change dramatically with diverging
derivatives.

\subsection{Observables}

Another theoretical interest in non-Hermitian $\mathcal{PT}$-symmetric
system is that the unitary evolution can be obtained by introducing metric
operator. In this section, we will illustrate the basic ideas via the above
analytical solution.

By introducing $\eta $-metric operator inner product, $\langle \cdot |\cdot
\rangle _{\eta }=$ $\left\langle \cdot \right\vert \eta \left\vert \cdot
\right\rangle $, time evolution can be expressed in a unitary way and also a
fully consistent quantum theory can be established~\cite{AM37,Bender98}.
Accordingly, the physical observables $O$ with respect to the metric
operator $\eta $\ can be constructed to meet the relation~\cite{AM38}
\end{subequations}
\begin{equation}
\eta O\eta ^{-1}=O^{\dag }.  \label{AliM_obser}
\end{equation}%
We examine the total particle number operator. It is defined as
\begin{equation}
\hat{N}_{p}=\sum_{l=1}^{N}a_{l}^{\dagger }a_{l}=\sum_{k_{l}}\bar{a}%
_{k_{l}}a_{k_{l}},
\end{equation}%
In the invariant subspace $V^{N_{p}}$,\ we have $[\hat{N}_{p},H^{\text{free}%
}]=0$, which allows the eigen equations of the operators\ in the form of
\begin{equation}
\begin{aligned} \hat{N}_{p}\overline{\left\vert
n_{k_{1}},n_{k_{2}},...,n_{k_{N}}\right\rangle } &=N_{p}\overline{\left\vert
n_{k_{1}},n_{k_{2}},...,n_{k_{N}}\right\rangle }, \\ \hat{N}_{p}\left\vert
n_{k_{1}},n_{k_{2}},...,n_{k_{N}}\right\rangle &=N_{p}\left\vert
n_{k_{1}},n_{k_{2}},...,n_{k_{N}}\right\rangle , \end{aligned}
\label{N_on_eigenstates}
\end{equation}%
and%
\begin{equation}
\hat{N}_{p}\left\vert n_{_{1}},n_{_{2}},...,n_{_{N}}\right\rangle
=N_{p}\left\vert n_{_{1}},n_{_{2}},...,n_{_{N}}\right\rangle .
\end{equation}%
These indicate that the Hermitian operator $\hat{N}_{p}$ is an observable.
Alternatively, this can be proved in the framework of non-Hermitian quantum
mechanics. Actually, we have%
\begin{align}
\eta \hat{N}_{p}\eta ^{-1}&
=\sum_{\{n_{k_{i}}\}}|n_{k_{1}},n_{k_{2}},...,n_{k_{N}}\rangle \langle
n_{k_{1}},n_{k_{2}},...,n_{k_{N}}|  \label{Np_hat} \\
& \times \hat{N}_{p}\sum_{\{n_{k_{i}^{\prime }}\}}\overline{%
|n_{k_{1}^{\prime }},n_{k_{2}^{\prime }},...,n_{k_{N}^{\prime }}\rangle }%
\overline{\langle n_{k_{1}^{\prime }},n_{k_{2}^{\prime
}},...,n_{k_{N}^{\prime }}|}  \notag \\
& =N_{p}\sum_{\{n_{k_{i}}\}}\left\vert
n_{k_{1}},n_{k_{2}},...,n_{k_{N}}\right\rangle \overline{\langle
n_{k_{1}},n_{k_{2}},...,n_{k_{N}}|}  \notag \\
& =\hat{N}_{p}^{\dagger }  \notag
\end{align}%
since the biorthogonal basis satisfies%
\begin{equation}
\sum_{\{n_{k_{i}}\}}\left\vert
n_{k_{1}},n_{k_{2}},...,n_{k_{N}}\right\rangle \overline{\langle
n_{k_{1}},n_{k_{2}},...,n_{k_{N}}|}=1.  \label{multi_biorthogonal}
\end{equation}%
Accordingly, for operator $\bar{n}_{k_{l}}=\bar{a}_{k_{l}}a_{k_{l}}$, we
also have
\begin{align}
\eta \bar{n}_{k_{l}}\eta ^{-1}&
=\sum_{\{n_{k_{i}}\}}|n_{k_{1}},n_{k_{2}},...,n_{k_{N}}\rangle \langle
n_{k_{1}},n_{k_{2}},...,n_{k_{N}}|  \label{nkl_bar} \\
& \times \bar{n}_{k_{l}}\sum_{\{n_{k_{i}^{\prime }}\}}\overline{%
|n_{k_{1}^{\prime }},n_{k_{2}^{\prime }},...,n_{k_{N}^{\prime }}\rangle }%
\overline{\langle n_{k_{1}^{\prime }},n_{k_{2}^{\prime
}},...,n_{k_{N}^{\prime }}|}  \notag \\
& =\sum_{\{n_{k_{i}}\}}n_{k_{l}}\left\vert
n_{k_{1}},n_{k_{2}},...,n_{k_{N}}\right\rangle \overline{\langle
n_{k_{1}},n_{k_{2}},...,n_{k_{N}}|}  \notag \\
& =\bar{n}_{k_{l}}^{\dagger },  \notag
\end{align}%
in order to obtain (\ref{Np_hat}) and (\ref{nkl_bar}) we used (\ref%
{four_a_operators}), (\ref{N_on_eigenstates}), and (\ref{multi_biorthogonal}%
).

Then we conclude that the two types of particle number operators, $\hat{N}%
_{p}=\sum_{l=1}^{N}a_{l}^{\dagger }a_{l}$ and $\bar{n}_{k_{l}}=\bar{a}%
_{k_{l}}a_{k_{l}}$, are both observables. Besides, the Hamiltonian $H^{\text{%
\textrm{free}}}$ itself and the metric operator $\eta $ are also
observables. Here, we would like to clarify that $\hat{N}_{p}$ and $\eta $
are both Hermitian operators, but $H^{\text{\textrm{free}}}$ and $\bar{n}%
_{k_{l}}$ are non-Hermitian operators. However, operators $a_{i}^{\dagger
}a_{i}$\ and $a_{k_{l}}^{\dagger }a_{k_{l}}$\ are no longer observables~\cite%
{AM37,Heiss40,BenderRPP}, which are proved through a simple illustration in
the \ref{sec_appendix}.

Here we want to stress that, the term "observable" is specific to the
non-Hermitian quantum mechanics framework, which differs from that in
traditional quantum mechanics. It is still controversial for the
interpretation of the observable. As pointed above, $\hat{N}_{p}$ is a good
quantum number, or say, the obtained eigenstates of $H$ are also the
eigenstates of the total particle number $\hat{N}_{p}$. This guarantees a
unitary time evolution if the $\eta $-metric operator inner product is
taken. However, the Dirac expectation value of the particle number is not
conservative under time evolution, which seems to be expected. This is
basically caused by the fact that the eigenstates of $\mathcal{PT}$%
-symmetric Hamiltonian are non-orthogonal under the Dirac inner product. On
the other hand, all Hermitian operators are observables in Hermitian quantum
physics. For example, operator $a_{i}^{\dagger }a_{i}$\ is an observable in
Hermitian quantum mechanics but not regarded as an observable according to
the non-Hermitian theory. Nevertheless, it is worth to mention that the
observation of the non-Hermitian behavior in experiment, e.g., the power
oscillation phenomenon \cite{PTinOpticsPO}\cite{AGuo}, is based on the
distribution of Dirac expectation value for $a_{i}^{\dagger }a_{i}$.

\section{Nonzero interaction system}

\label{sec_nonzero_interaction}Now we turn to investigate scaling behavior
and phase diagram of the system at nonzero $U$. The boundary of the phase is
the main character for a non-Hermitian system. So far most studies dealt
with the noninteracting system. For a non-Hermitian lattice model, it is
shown that the critical point is sensitive to the distribution of the
coupling constant and on-site potential~\cite{Giorgi,Bendix,Joglekar}. It
indicates that the inhomogeneity of a noninteracting system may shrink the\
unbroken region of $\mathcal{PT}$\ symmetry. From the point of view of mean
field theory, on-site interaction takes the role of on-site potentials in
some sense. Thus it is presumable that a nonzero $U$ may shift the critical
point. In most cases of nonzero $U$, an exact solution is hard to obtain. In
this paper, we only consider the two-particle case within some specific
parameter areas.

\subsection{Solutions for nonzero $U$}

The two-particle Bethe ansatz solution
\begin{equation}
\left\vert \psi _{k_{1},k_{2}}\right\rangle
=\sum_{l_{1},l_{2}}f_{k_{1},k_{2}}\left( l_{1},l_{2}\right)
a_{l_{1}}^{\dagger }a_{l_{2}}^{\dagger }\left\vert \text{vac}\right\rangle
\end{equation}%
where the explicit form of $f_{k_{1},k_{2}}\left( l_{1},l_{2}\right) $ can
be expressed as
\begin{gather}
f_{k_{1},k_{2}}\left( l_{1},l_{2}\right) =A\left( k_{1},k_{2}\right)
e^{ik_{1}l_{1}+ik_{2}l_{2}}+A\left( k_{2},k_{1}\right)
e^{ik_{2}l_{1}+ik_{1}l_{2}}+A\left( k_{1},-k_{2}\right)
e^{ik_{1}l_{1}-ik_{2}l_{2}}+A\left( k_{2},-k_{1}\right)
e^{ik_{2}l_{1}-ik_{1}l_{2}}  \label{fk_long} \\
+A\left( -k_{1},k_{2}\right) e^{-ik_{1}l_{1}+ik_{2}l_{2}}+A\left(
-k_{2},k_{1}\right) e^{-ik_{2}l_{1}+ik_{1}l_{2}}+A\left(
-k_{1},-k_{2}\right) e^{-ik_{1}l_{1}-ik_{2}l_{2}}+A\left(
-k_{2},-k_{1}\right) e^{-ik_{2}l_{1}-ik_{1}l_{2}}  \notag
\end{gather}%
Based on the stationary Schr\"{o}dinger equation%
\begin{equation}
H\left\vert \psi _{k_{1},k_{2}}\right\rangle =E(k_{1},k_{2})\left\vert \psi
_{k_{1},k_{2}}\right\rangle ,  \label{Schrodinger_Eq}
\end{equation}%
quasimomenta $k_{1}$ and $k_{2}$ satisfy the equations%
\begin{gather}
\frac{\left( J^{2}+\gamma ^{2}e^{i2k_{1}}\right) e^{-ik_{1}\left( N+1\right)
}}{\left( J^{2}+\gamma ^{2}e^{-i2k_{1}}\right) e^{ik_{1}\left( N+1\right) }}=%
\frac{G_{-+}G_{++}}{G_{--}G_{+-}},  \label{k1} \\
k_{1}\leftrightarrow k_{2}.  \label{k2}
\end{gather}%
where
\begin{equation}
G_{\sigma \sigma ^{\prime }}=2Ji\sin k_{1}+\sigma 2Ji\sin k_{2}+\sigma
^{\prime }U
\end{equation}%
with $\sigma $, $\sigma ^{\prime }=\pm $. Hereafter $k_{1}\leftrightarrow
k_{2}$\ denotes the corresponding equation by exchanging $k_{1}$ and $k_{2}$%
.\ The quasimomenta $k_{1}$, $k_{2}$ and amplitudes $A\left(
k_{1},k_{2}\right) $ can be determined by (\ref{Schrodinger_Eq}) and the
proper definition of inner product according to the $\mathcal{PT}$-symmetric
quantum theory. The corresponding eigenvalue is $E(k_{1},k_{2})=-2J\left(
\cos k_{1}+\cos k_{2}\right) ,$ the reality of $E(k_{1},k_{2})$ determines
the phase diagram of the system.\ It is obvious that (\ref{k1}) and (\ref{k2}%
) are invariant under $k_{1}\rightarrow k_{2}$, $k_{2}\rightarrow k_{1}$;
and also under $k_{1}\rightarrow -k_{1}$, $k_{2}\rightarrow -k_{2}$. We
rewritten (\ref{k1}) and (\ref{k2}) explicitly as%
\begin{gather}
\begin{array}{r}
\left( U/J\right) \sin k_{1}\left\{ \cos \left[ k_{1}\left( N+1\right) %
\right] +\left( \gamma /J\right) ^{2}\cos \left[ k_{1}\left( N-1\right) %
\right] \right\} \\
+\left\{ \sin \left[ k_{1}\left( N+1\right) \right] +\left( \gamma /J\right)
^{2}\sin \left[ k_{1}\left( N-1\right) \right] \right\} \\
\times \left[ \sin ^{2}k_{2}-\sin ^{2}k_{1}+\left( U/2J\right) ^{2}\right] =0%
\end{array}%
,  \label{k1k2} \\
k_{1}\leftrightarrow k_{2}.  \label{k2k1}
\end{gather}%
Although the analytical solutions of (\ref{k1k2}) and (\ref{k2k1}) are hard
to obtain, approximate solutions within certain ranges of the parameters $%
\gamma $, $J$,\ and $U$ may shed light on the influence of $U$ on the phase
boundary.

\subsection{Solutions at the point $\protect\gamma =J$}

We start with the solution of an even $N$ system at the point $\gamma =J$,
which is the exceptional point for the system of $U=0$. The influence of
on-site interaction on the phase boundary can be qualitatively revealed.
Taking $\gamma =J$, (\ref{k1k2}) and (\ref{k2k1}) are reduced to%
\begin{gather}
U\sin k_{1}\cos \left( k_{1}N\right) +J\left[ \left( U/2J\right) ^{2}+\cos
^{2}k_{1}\right] \times \sin \left( k_{1}N\right) =0,k_{2}=\pi /2,
\label{reduced
Eq} \\
k_{1}\leftrightarrow k_{2}.
\end{gather}%
respectively. We notice that $k_{1}=\pi /2$ is the solution of (\ref{reduced
Eq}) for $U=0$. Then for small $U$ case, the solutions should have the form $%
(\pi /2,\pi /2+\theta _{2})$ and $(\pi /2+\theta _{1},\pi /2)$. For
sufficient small $U$, taking the approximations $\sin \left( \theta
_{1,2}N\right) \approx \theta _{1,2}N$ and $\cos \left( \theta
_{1,2}N\right) \approx 1$, the critical equation reduces to%
\begin{equation}
\theta _{1,2}^{3}+\left( U/2J\right) ^{2}\theta _{1,2}+U/\left( JN\right) =0
\label{app eq}
\end{equation}%
which has one real root and two non-real complex conjugate roots, since the
discriminant of the cubic equation $\Delta =\left[ U/\left( 2JN\right) %
\right] ^{2}$ $+[\left( U/2J\right) ^{2}/3]^{3}>0$. Furthermore, one can get
the solution of the cubic equation by routine. In order to obtain a concise
expression of $\theta _{1,2}$, we simply ignore the term of $U^{2}$ in the
cubic equation, and then obtain $\theta _{1,2}=-\sqrt[3]{U/JN}$, $(1\pm i%
\sqrt{3})\sqrt[3]{U/8JN}$. The corresponding complex conjugate eigenvalues
are $E_{\pm }=2J\sin \theta _{1,2}$ $\approx (1\pm i\sqrt{3})\sqrt[3]{%
UJ^{2}/N}$. Then we can conclude that point $\gamma =J$ is in the broken $%
\mathcal{PT}$-symmetric region in the presence of on-site interaction, which
shrinks the unbroken\ region of $\mathcal{PT}$ symmetry. Note that for a
given $N$, the eigenvalues $E_{\pm }$ become further away from real values
as $U$ grows. This agrees with the numerical simulation for phase diagram of
the finite size systems.

\subsection{Exceptional points and scaling behavior}

Now we focus on the phase boundary of the system with small $U$. It is
presumable that one pair of coalescing eigenstates have the quasimomenta $%
k_{1,2}=$ $\pi /2+\delta _{1,2}$\ with $\left\vert \delta _{1,2}\right\vert
\ll 1$. The original critical (\ref{k1k2}) and (\ref{k2k1}) are reduced to%
\begin{gather}
\begin{array}{r}
\left( U/J\right) \cos \delta _{1}\left\{ \sin \left[ \delta _{1}\left(
N+1\right) \right] -\left( \gamma /J\right) ^{2}\sin \left[ \delta
_{1}\left( N-1\right) \right] \right\} \\
-\left\{ \cos \left[ \delta _{1}\left( N+1\right) \right] -\left( \gamma
/J\right) ^{2}\cos \left[ \delta _{1}\left( N-1\right) \right] \right\} \\
\times \left[ \cos ^{2}\delta _{2}-\cos ^{2}\delta _{1}+\left( U/2J\right)
^{2}\right] =0%
\end{array}%
,  \label{delta1delta2} \\
\delta _{2}\leftrightarrow \delta _{1}.  \label{delta2delta1}
\end{gather}%
Furthermore, under the approximation $\left\vert \delta _{1,2}\right\vert
N\ll 1$ and ignoring the term of $U^{2}$, (\ref{k1k2}) and (\ref{k2k1}) are
reduced to polynomial equations
\begin{gather}
\left( \delta _{1}^{2}-\delta _{2}^{2}\right) \left( \zeta -\delta
_{1}^{2}\right) -\left( N^{2}\zeta +1\right) \delta _{1}u=0,
\label{RCE_stage_2_a} \\
\delta _{2}\leftrightarrow \delta _{1},  \label{RCE_stage_2_b}
\end{gather}%
where we have defined
\begin{equation}
\zeta N=\frac{J^{2}-\gamma ^{2}}{J^{2}+\gamma ^{2}},uN=\frac{U}{J}.
\end{equation}%
Eliminating $\delta _{2}$, one can obtain the equation about $\delta _{1}$
in the form of
\begin{equation}
f\left( \delta _{1}\right) =\delta _{1}^{6}-\zeta \delta _{1}^{4}+u\left(
\zeta N^{2}+1\right) \delta _{1}^{3}-\zeta ^{2}\delta _{1}^{2}+\zeta ^{3}=0,
\label{f_delta1}
\end{equation}%
which solution determines the\ eigenvalues. As pointed out in Ref.~\cite%
{JLPT}, when the eigenstates turn to coalescence at the critical point $%
\gamma _{c}$, $f\left( \delta _{1}\right) $\ should also satisfy the
equation d$f\left( \delta _{1}\right) /$d$\delta _{1}=0$. Eliminating $%
\delta _{1}$ from (\ref{f_delta1}) and d$f\left( \delta _{1}\right) /$d$%
\delta _{1}=0$, we have
\begin{equation}
3^{3}u^{2}\left( \zeta N^{2}+1\right) ^{2}-2^{8}\zeta ^{3}=0  \label{zeta_N}
\end{equation}%
under the condition $\left\vert \zeta \right\vert N^{2}\ll 1$. Then we can
obtain $\gamma _{c}$\ approximately as
\begin{figure}[tb]
\includegraphics[ bb=38 158 548 630, width=5.5 cm, clip]{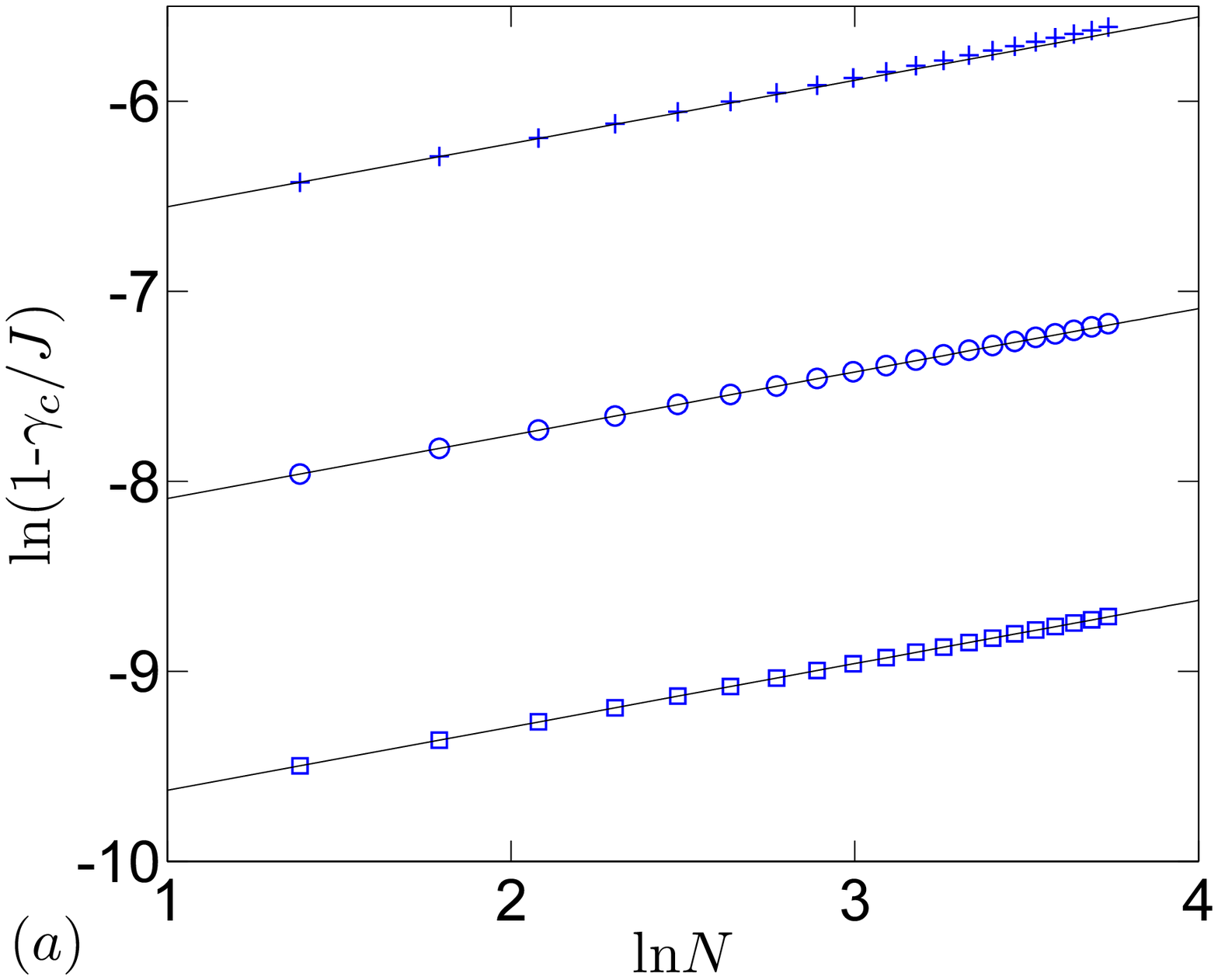} %
\includegraphics[ bb=38 158 548 630, width=5.5 cm, clip]{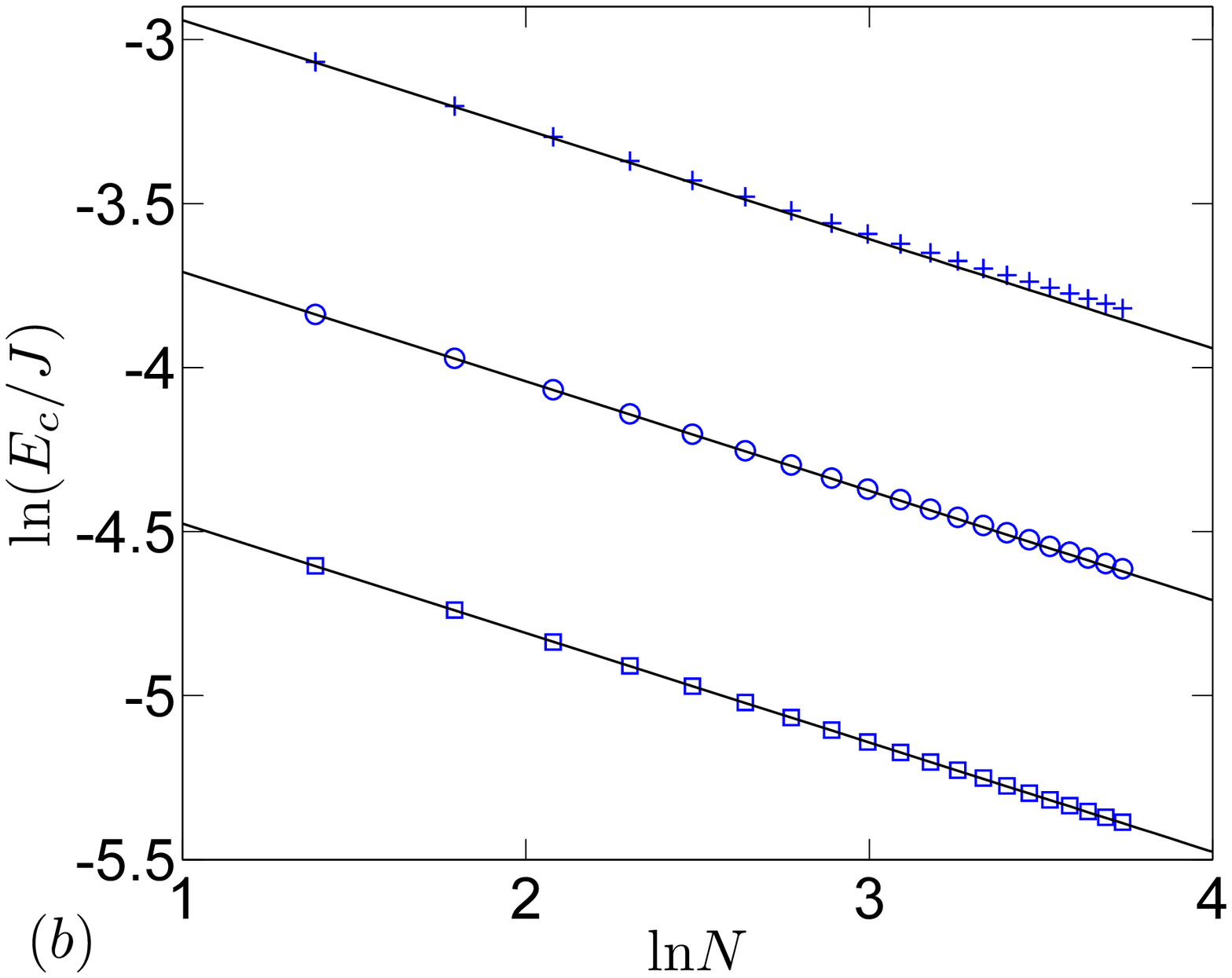} %
\includegraphics[ bb=46 197 500 610, width=5.5 cm, clip]{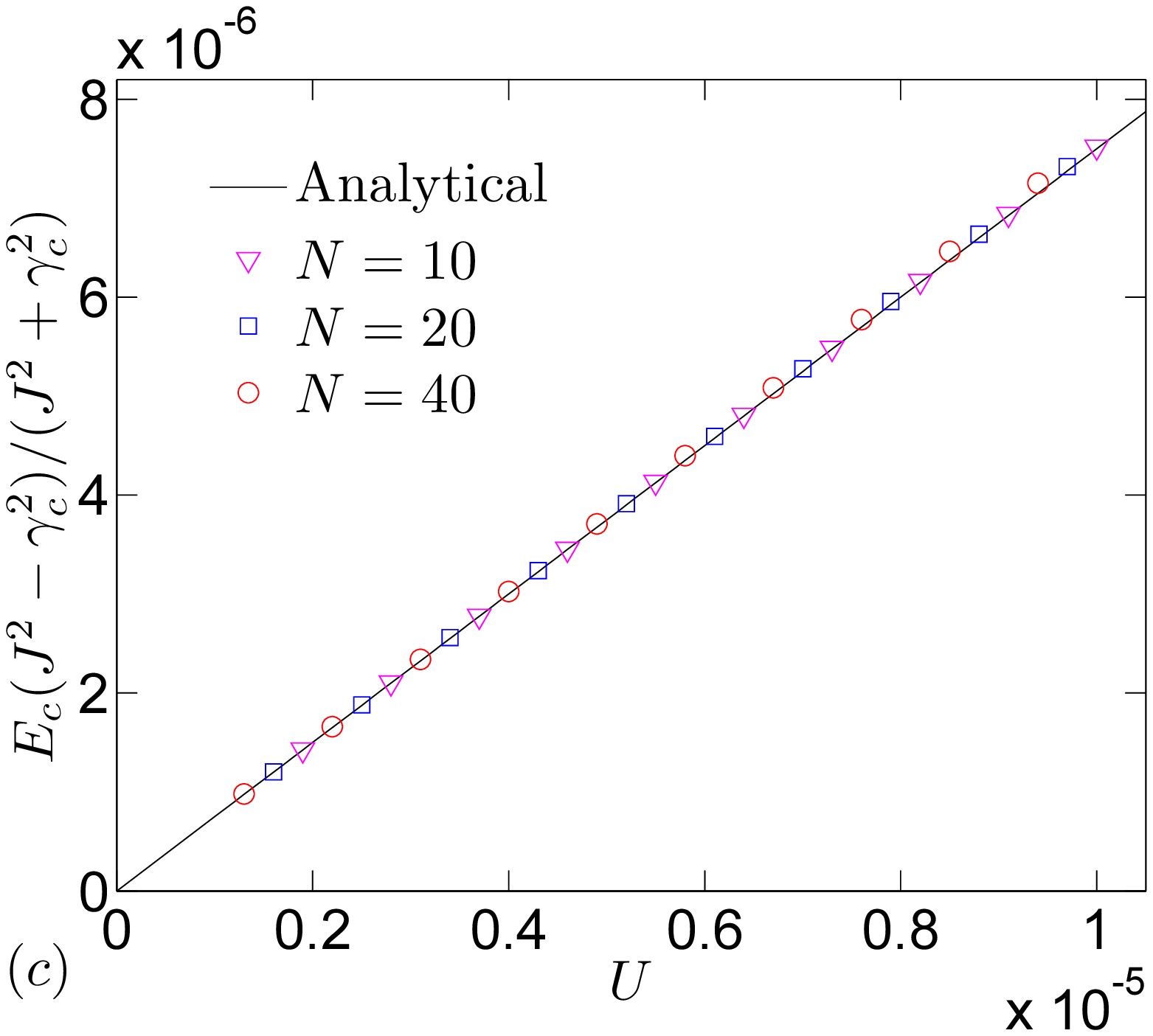}
\caption{(Color online) Plots of (\protect\ref{gamma_N}), (\protect\ref%
{E_c_N}), (\protect\ref{scaling}) and the corresponding numerical simulation
obtained by exact diagonalization. In (a) and (b), the blue crosses, circles
and squares indicate the numerical results for the cases of $U=10^{-4}$, $%
10^{-5}$ and $10^{-6}$, respectively. The black lines are the plots of the
corresponding analytical expressions. It shows that they are in agreement
with each other for $U=10^{-5}$ and $10^{-6}$. In the case of $U=10^{-4}$, a
slight deviation appears for large $N$. From (c), it is observed that the
numerical result accords with the analytical expression very well for $N=10$
and $20$. A slight deviation appears for large $U$ in the $N=40$ system.}
\label{fig_ln}
\end{figure}
\begin{figure}[tbh]
\includegraphics[bb=40 178 547 590, width=5.5 cm,
clip]{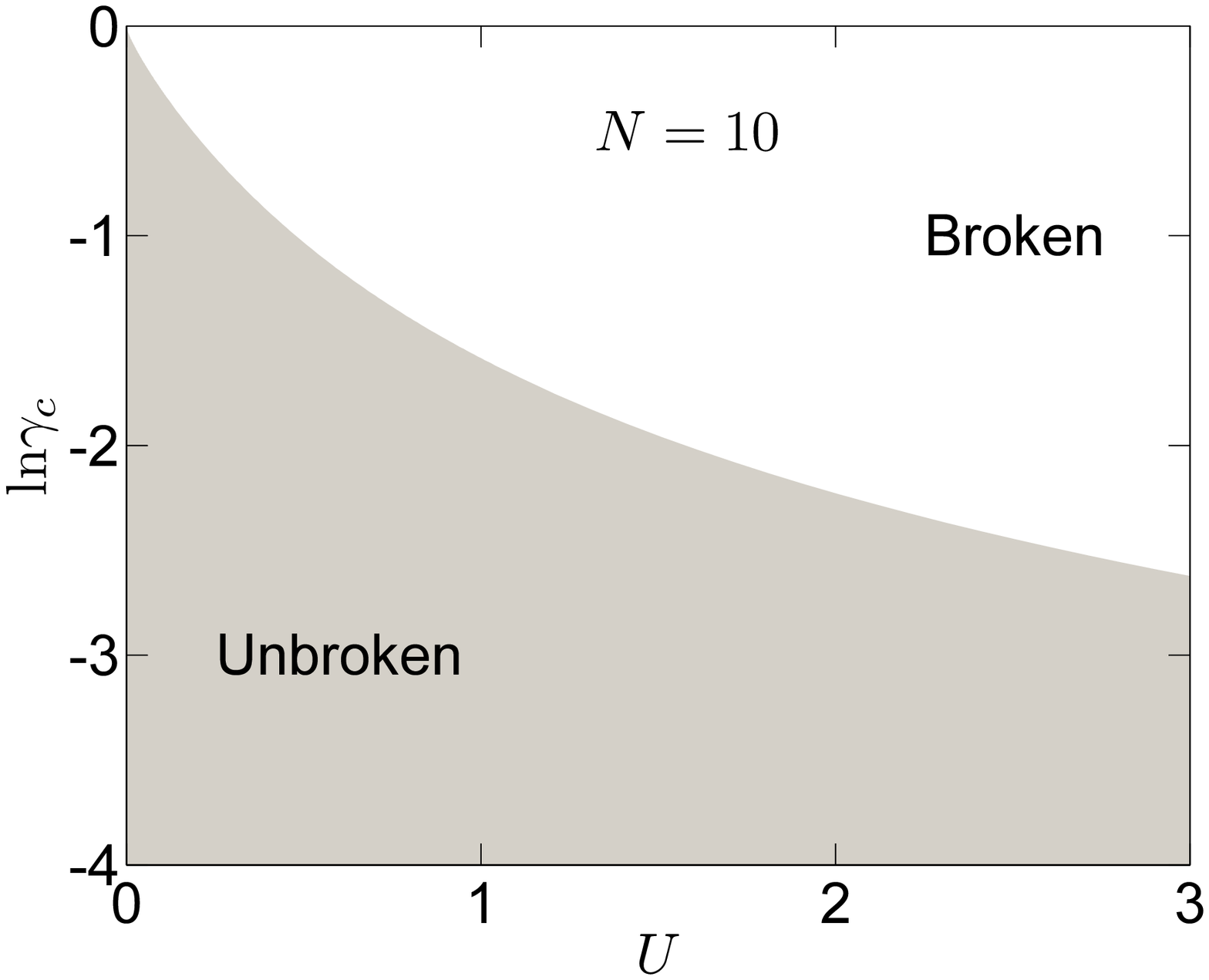}
\includegraphics[bb=40 178 547 590, width=5.5 cm,
clip]{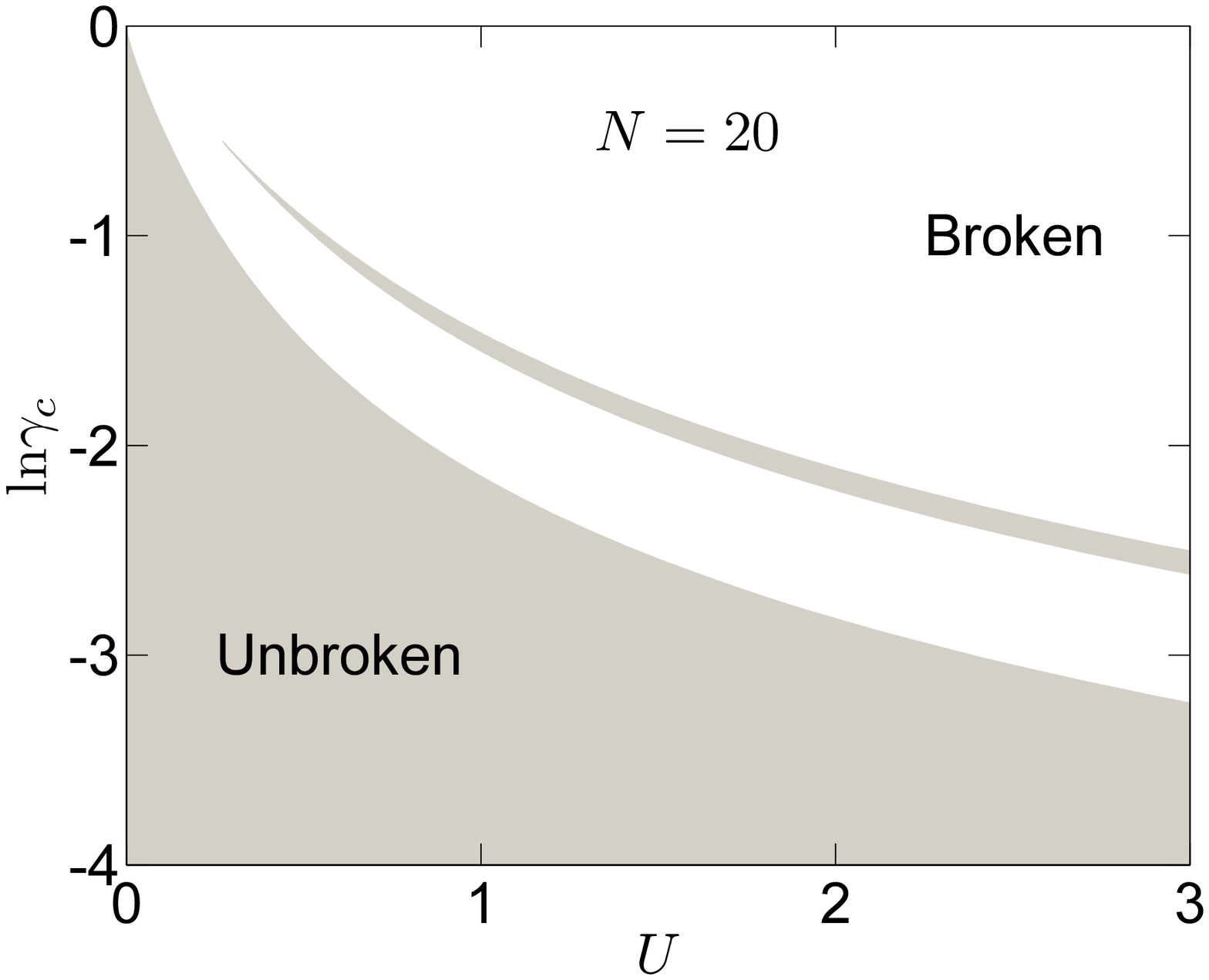}
\includegraphics[bb=40 178 547 590, width=5.5 cm,
clip]{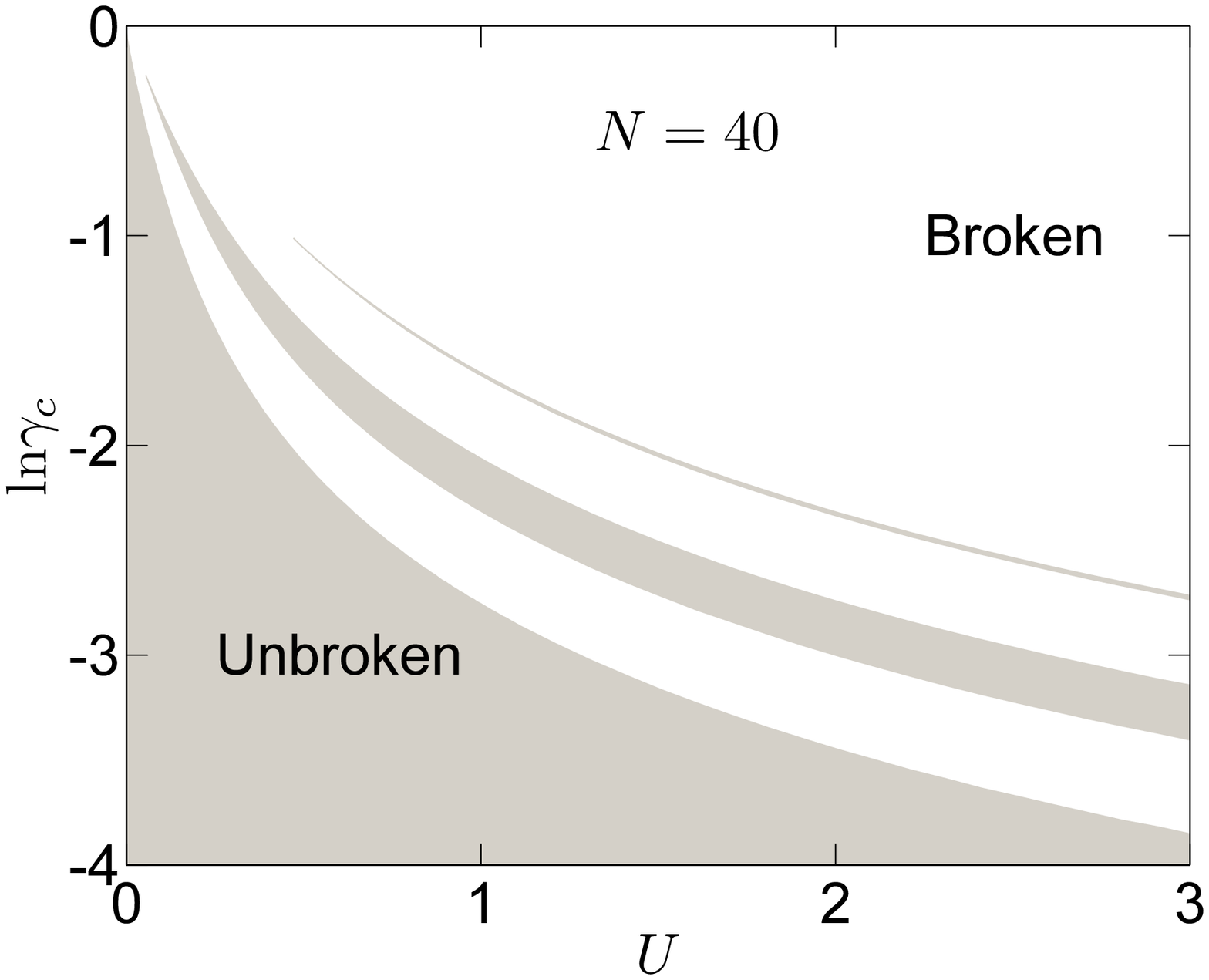}
\caption{Phase diagrams obtained by exact diagonalization for finite systems
with $N=10$, $20$, and $40$. The shadow area indicates the region where the
spectrum is entirely real. It shows that as $N$ increases the unbroken $%
\mathcal{PT}$-symmetric region becomes narrow. There are multiple unbroken $%
\mathcal{PT}$-symmetric regions appear as $N$ increases.}
\label{phase_diagram}
\end{figure}
\begin{equation}
\gamma _{c}=J\sqrt{\frac{1-\beta }{1+\beta }}\simeq J\left( 1-\beta \right) ,
\label{gamma_c}
\end{equation}%
where $\beta =\left( 3/8\right) \sqrt[3]{2NU^{2}/J^{2}}$. It shows that the
exceptional point exhibits an interaction sensitivity and undergoes dramatic
changes following the change of the Hubbard interaction. Substituting $%
\gamma _{c}$ into (\ref{RCE_stage_2_a}) and (\ref{RCE_stage_2_b}), we have
\begin{equation}
\delta _{1,2}=\frac{1\pm i\sqrt{2}}{2}\sqrt[3]{\frac{U}{2JN}},
\end{equation}%
which leads to the critical eigenvalue%
\begin{equation}
E_{c}=2J\left( \sin \delta _{1}+\sin \delta _{2}\right) \approx 2J\sqrt[3]{%
\frac{U}{2JN}}.  \label{E_c}
\end{equation}%
At $U=0$, (\ref{gamma_c}) and (\ref{E_c}) reproduce the obtained results in
our previous work~\cite{JLPT}: $\gamma _{c}=J$\ and $E_{c}=0$, respectively.
Furthermore, it is observed that for small $U$ and finite $N$, the critical
quantities $\gamma _{c}$\ and\ $E_{c}$\ can be expressed as
\begin{eqnarray}
\ln \left( 1-\gamma _{c}/J\right) &\approx &\frac{1}{3}\ln N+\ln [\frac{3}{%
2^{8/3}}\left( \frac{U}{J}\right) ^{2/3}],  \label{gamma_N} \\
\ln \left( E_{c}/J\right) &\approx &-\frac{1}{3}\ln N+\ln [2^{2/3}\left(
\frac{U}{J}\right) ^{1/3}],  \label{E_c_N}
\end{eqnarray}%
which shows the similar dependence on the size of the system. According to
the finite-size scaling ansatz~\cite{scaling book}, the critical behavior
can be extracted from the above-mentioned finite samples. Then combining (%
\ref{gamma_c}) with (\ref{E_c}) leads to
\begin{equation}
E_{c}\frac{J^{2}-\gamma _{c}^{2}}{J^{2}+\gamma _{c}^{2}}=\frac{3}{4}U,
\label{scaling}
\end{equation}%
which is a universal scaling law for such a phase transition in small $U$
limit. To verify and demonstrate the above analysis, numerical simulations
are performed to investigate the scaling behavior. We compute the quantities
$\gamma _{c}$ and $E_{c}$ for finite systems, which are plotted in Fig. \ref%
{fig_ln} as comparison with the analytical results (\ref{gamma_N}), (\ref%
{E_c_N}), and (\ref{scaling}). It shows that for small $U$, they are in
agreement with each other. It is worthy to note that the analytical
expressions in (\ref{gamma_N}), (\ref{E_c_N}), and (\ref{scaling}) are
obtained under the condition $UN^{2}/J\ll 2^{4}/\sqrt{3^{3}}$ (obtained from
$\left\vert \delta _{1,2}\right\vert N\ll 1$ and $\left\vert \zeta
\right\vert N^{2}\ll 1$). Thus for large size system, the scaling law holds
only within a very small parameter region. Nevertheless, our finding reveals
the fact that there should exist a universal scaling law for such kind of
phase transition.

In the interaction-free case, from Section \ref{sec_interaction_free} we
notice that the $\mathcal{PT}$ symmetry phase hardly changes as the system
size $N$ increasing. In contrast, for medium interaction $U$, we investigate
the phase diagram for finite size system by numerical simulation. The phase
boundary is determined from the reality of the eigenvalues obtained by exact
diagonalization. In Fig. \ref{phase_diagram}, the phase diagrams are plotted
as $\gamma _{c}$ versus $U$ for finite size chains. It shows that the
on-site interaction breaks the $\mathcal{PT}$ symmetry drastically.
Interestingly, there exist several $\mathcal{PT}$-symmetric regions and the
number of such regions increases as $N$ increases. The phase diagram shows
rich structure in cases of medium $U$.

\subsection{Phase transition induced by bound-pair state}

\begin{figure}[tbp]
\includegraphics[ bb=20 185 550 590, width=5.5 cm, clip]{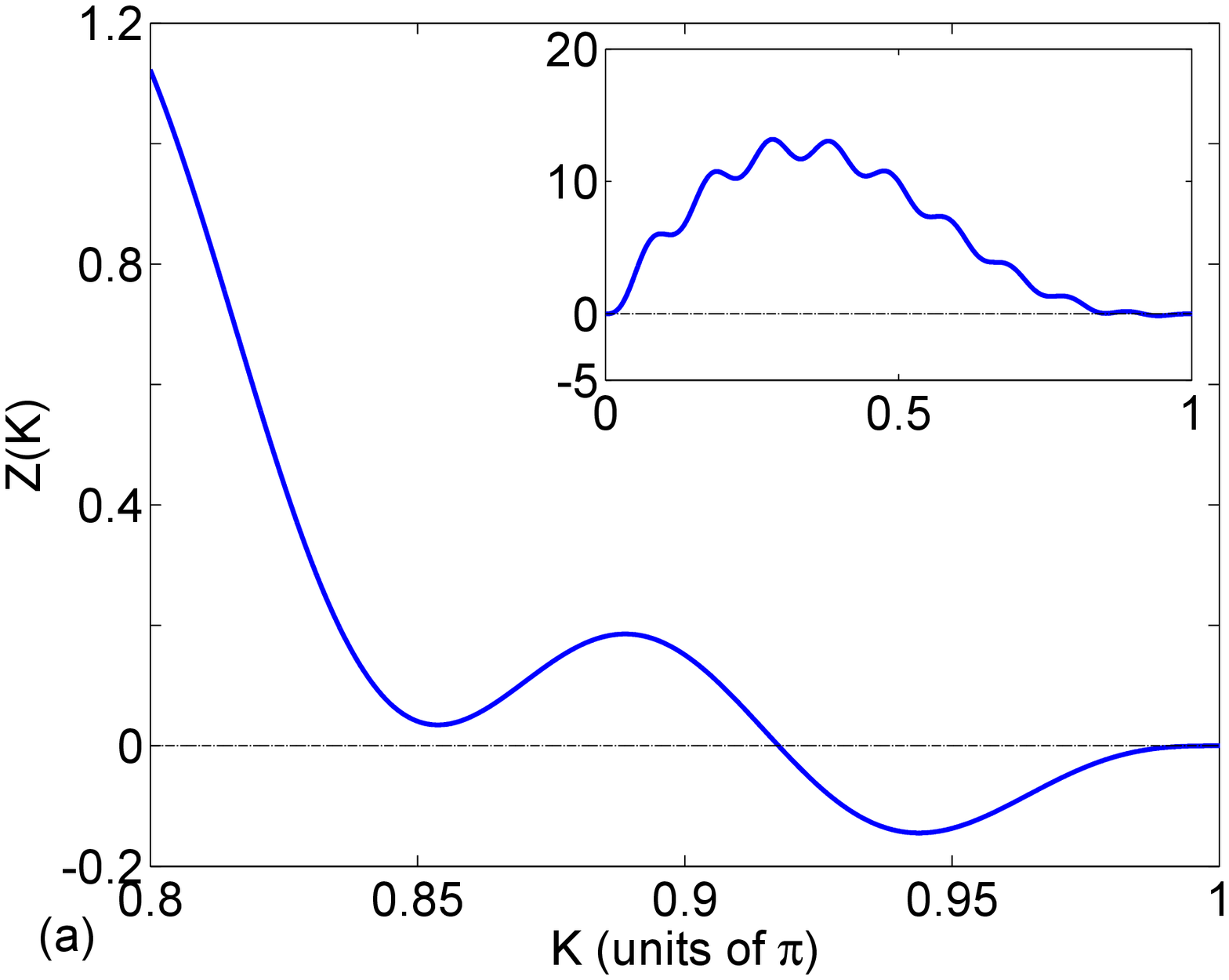} %
\includegraphics[ bb=20 185 550 590, width=5.5 cm, clip]{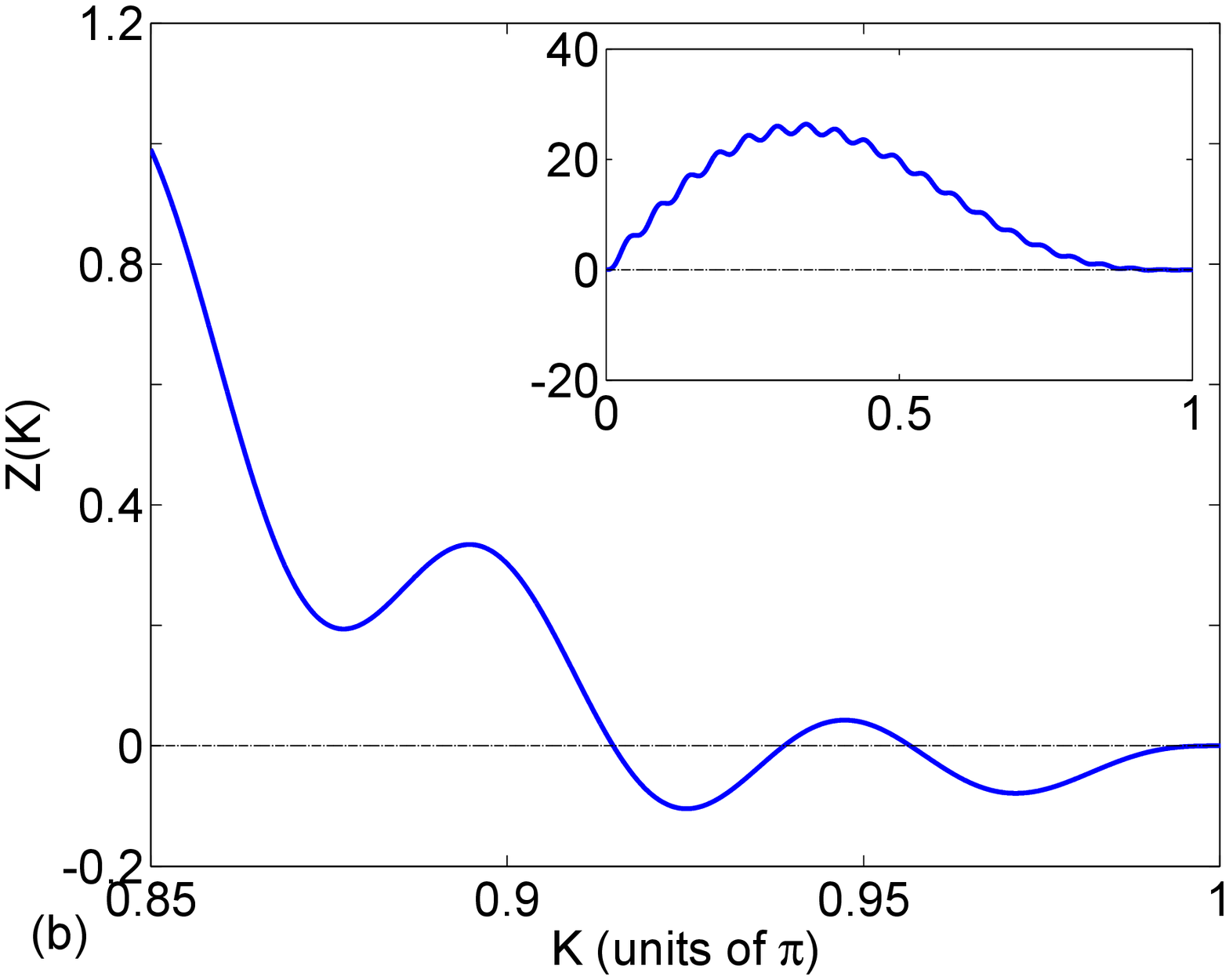} %
\includegraphics[ bb=20 185 550 590, width=5.5 cm, clip]{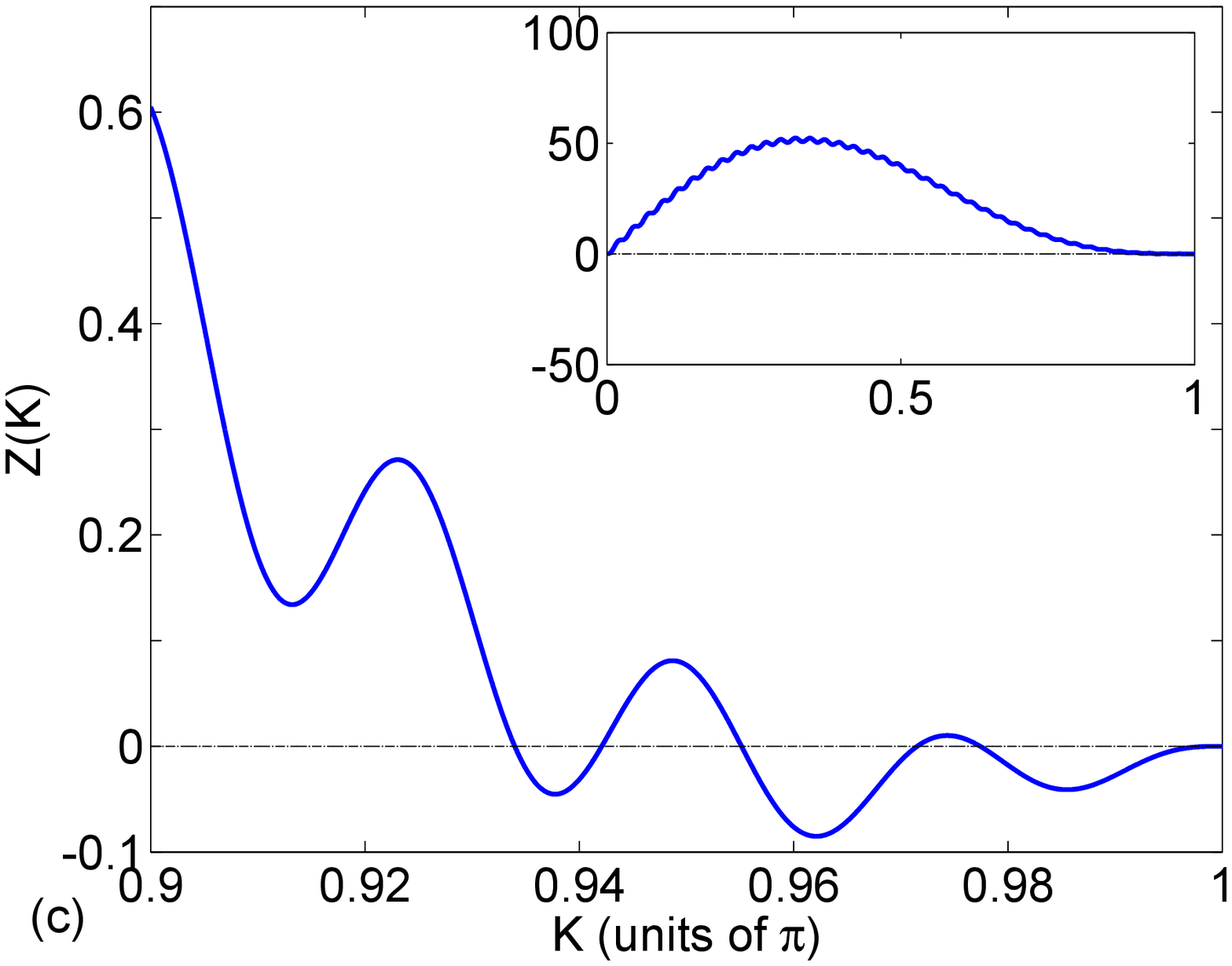} %
\includegraphics[ bb=20 185 550 590, width=5.5 cm, clip]{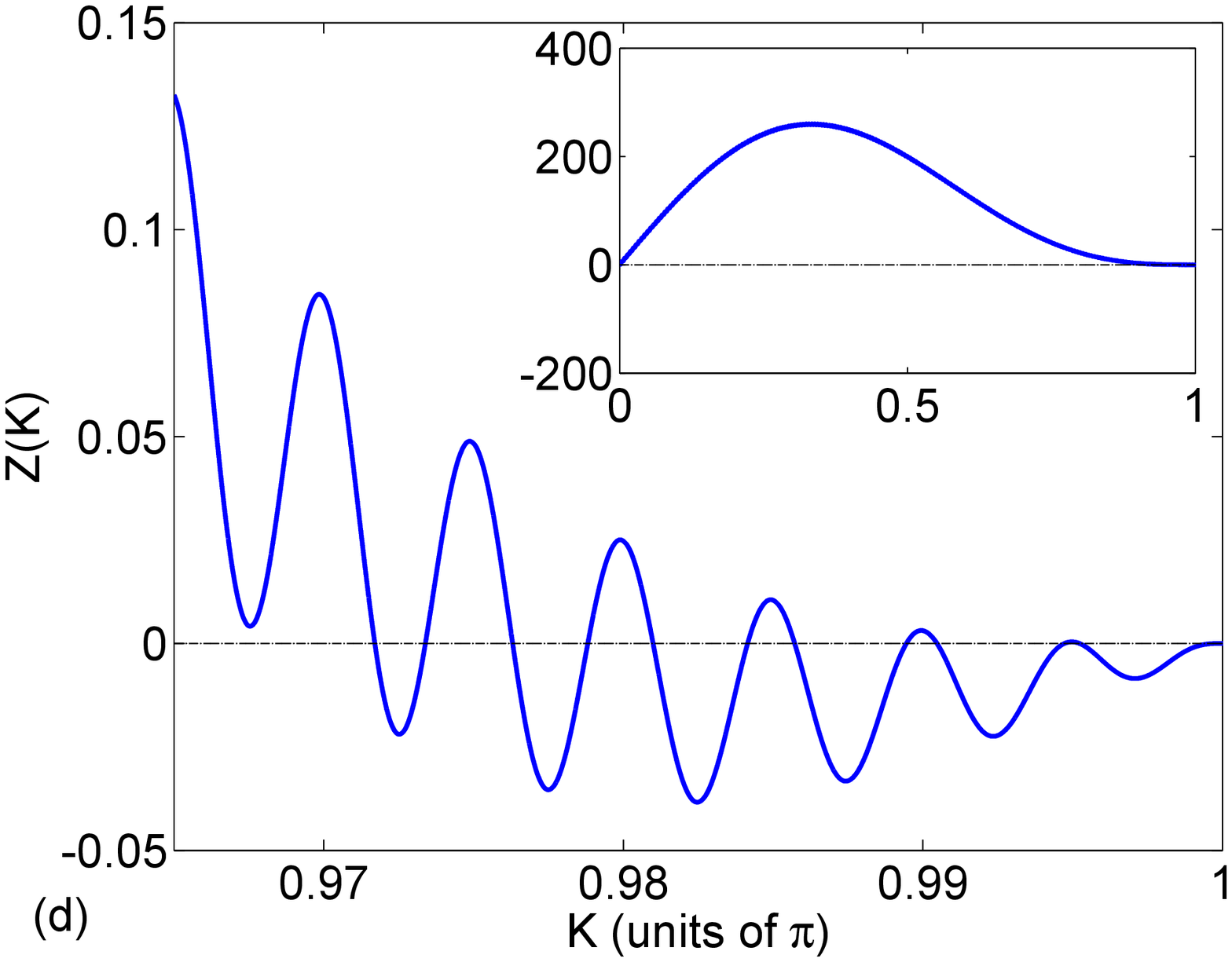} %
\includegraphics[ bb=20 185 550 590, width=5.5 cm, clip]{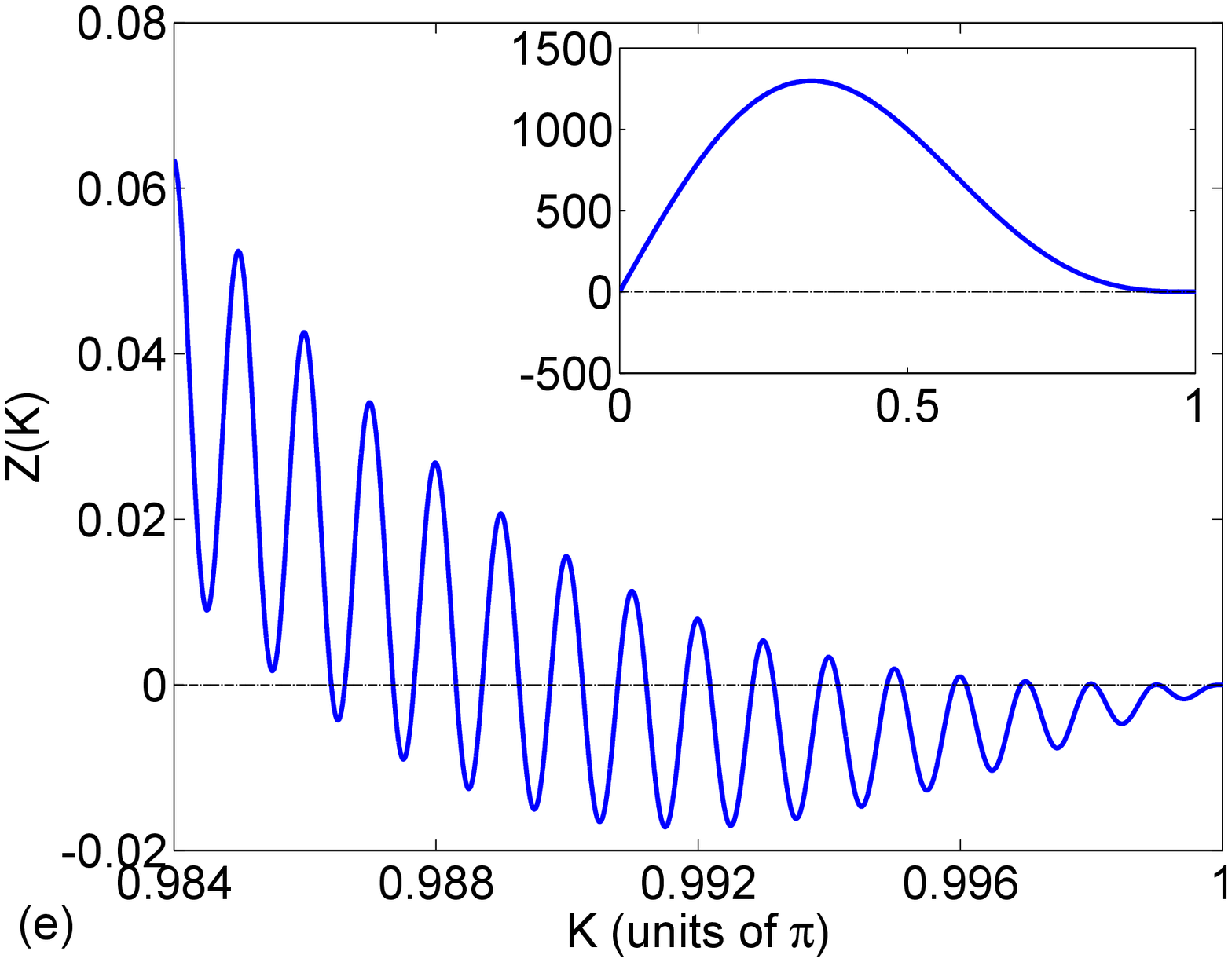} %
\includegraphics[ bb=20 185 550 590, width=5.5 cm, clip]{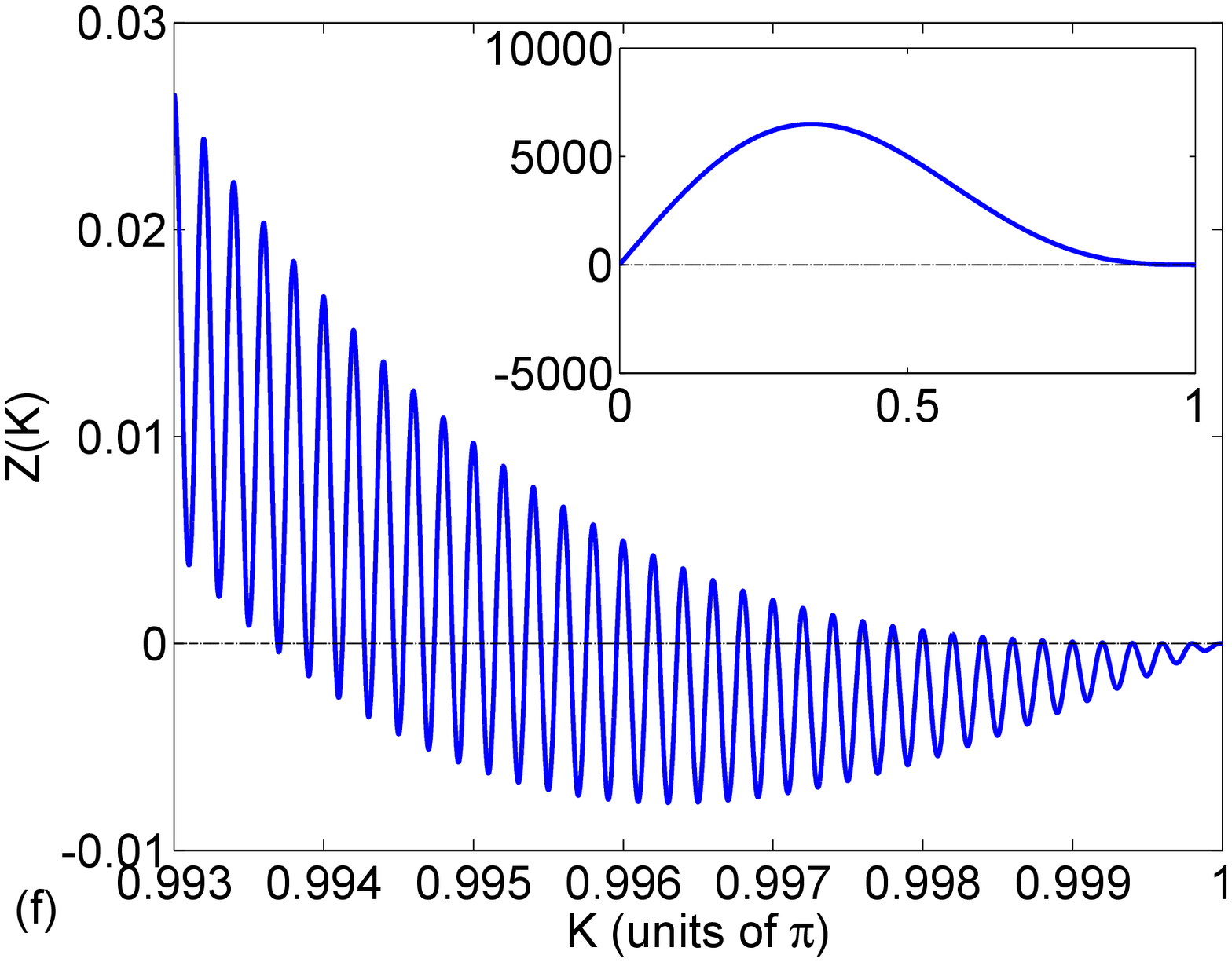}
\caption{Plots of function $Z\left( K\right) $, insert is $Z\left( K\right) $
in the region $\left( 0,\protect\pi \right) $. (a) $N=10$, (b) $N=20$, (c) $%
N=40$, (d) $N=200$, (e) $N=1000$, (f) $N=5000$. The number of roots of $%
Z(K)=0$ is $1, 3, 5, 11, 27, 63$, the number of unbroken $\mathcal{PT}$
symmetric regions is $1, 2, 3, 6, 14, 32$, respectively. The black line, $%
Z(K)$ being zero, is a guide to the eye only.}
\label{fig_Zk}
\end{figure}
In the following, we analytically investigate the boundaries of the $%
\mathcal{PT}$-symmetric phase of $H$. In strong on-site interaction case,
there exists bound-pair band induced by the interaction $U~$\cite%
{Winkler,JLBP,JLPerfetBP}.\ In the presence of $\gamma $, the $\mathcal{PT}$
symmetry is fragile. Here we focus on $\mathcal{PT}$-symmetric breaking
phase transition caused by the bound-pair state. We analyse the phase
diagram by introducing an effective Hamiltonian $H_{\mathrm{eff}}$, which
describes the bound states of system $H$ in strong on-site interaction $U$\
region. Based on the perturbation methods \cite{JLPerfetBP}, the effective
Hamiltonian $H_{\mathrm{eff}}$ reads%
\begin{equation}
H_{\mathrm{eff}}=\frac{2J^{2}}{U}\left( b_{i}^{\dagger }b_{i+1}+\mathrm{h.c.}%
\right) +\left( U+\frac{4J^{2}}{U}\right) b_{i}^{\dagger }b_{i}-\frac{2J^{2}%
}{U}\left( b_{1}^{\dagger }b_{1}+b_{N}^{\dagger }b_{N}\right) +2i\gamma
\left( b_{1}^{\dagger }b_{1}-b_{N}^{\dagger }b_{N}\right) ,  \label{H_eff}
\end{equation}%
where $b_{i}^{\dagger }=a_{i}^{\dagger 2}/\sqrt{2}$ ($b_{i}=a_{i}^{2}/\sqrt{2%
}$) is the bound pair creation (annihilation) operator on site $i$.

\begin{figure}[tbp]
\centering
\includegraphics[ bb=40 175 570 600, width=5.5 cm, clip]{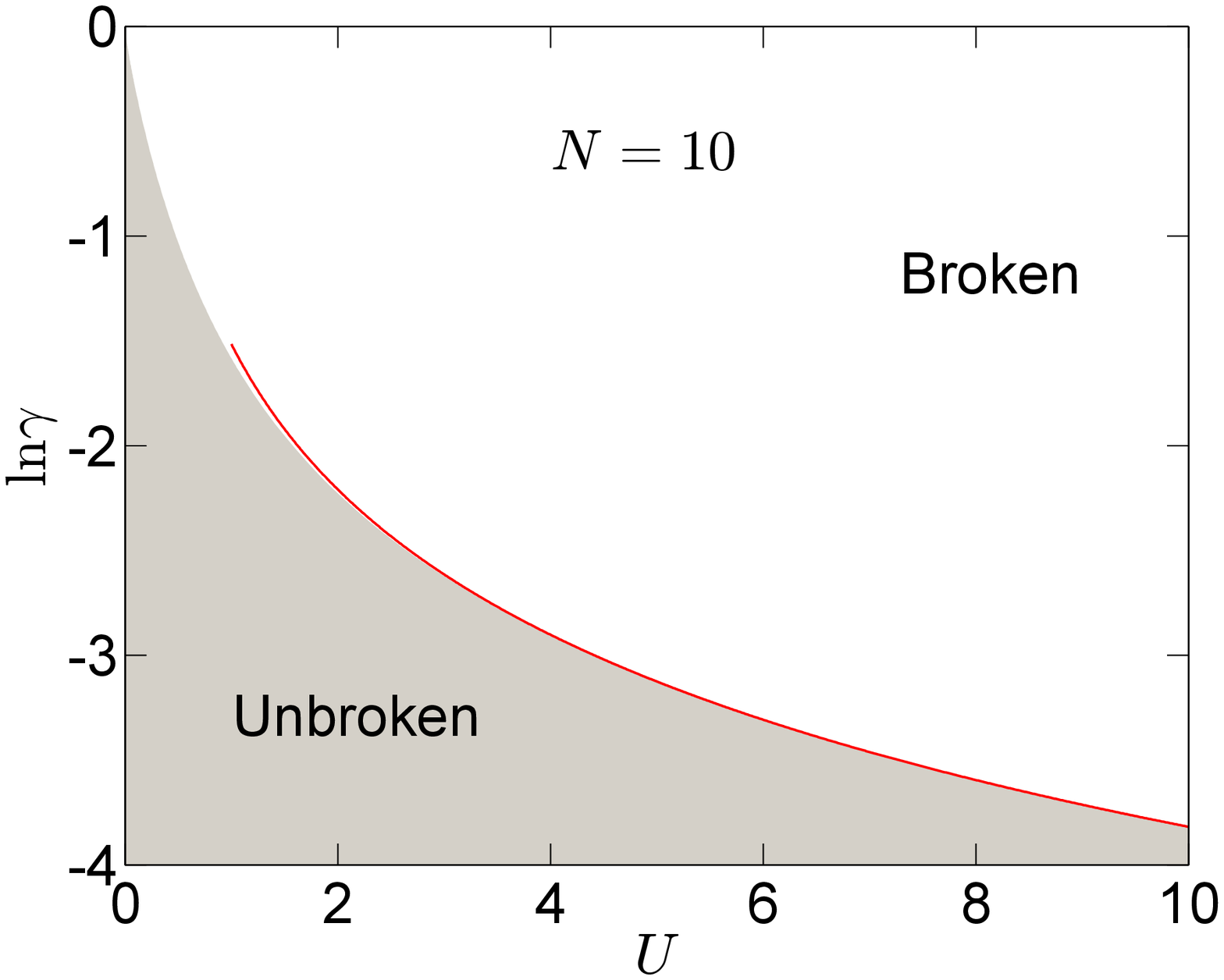}%
\includegraphics[ bb=40 175 570 600, width=5.5 cm, clip]{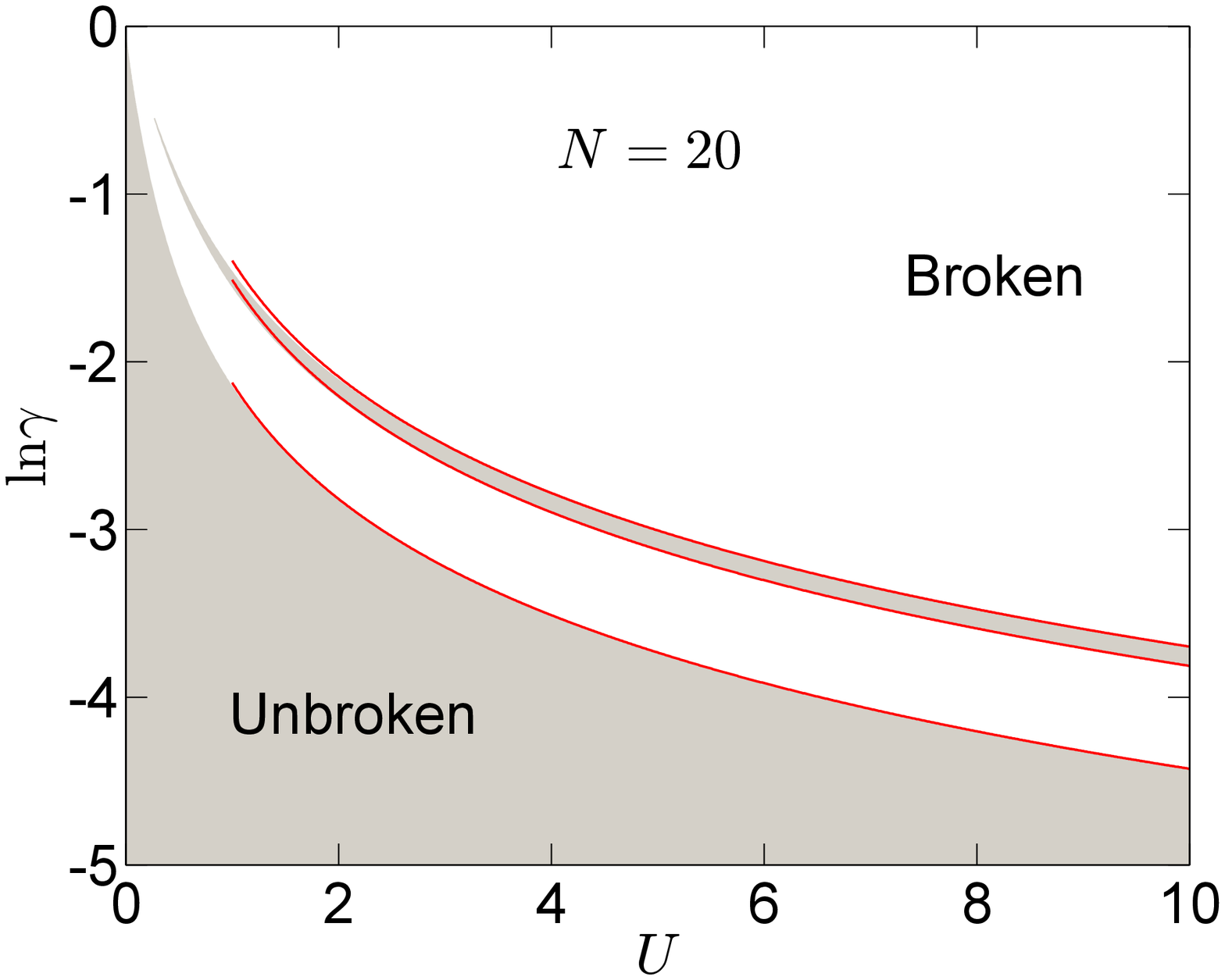}%
\includegraphics[ bb=40 175 570 600, width=5.5 cm, clip]{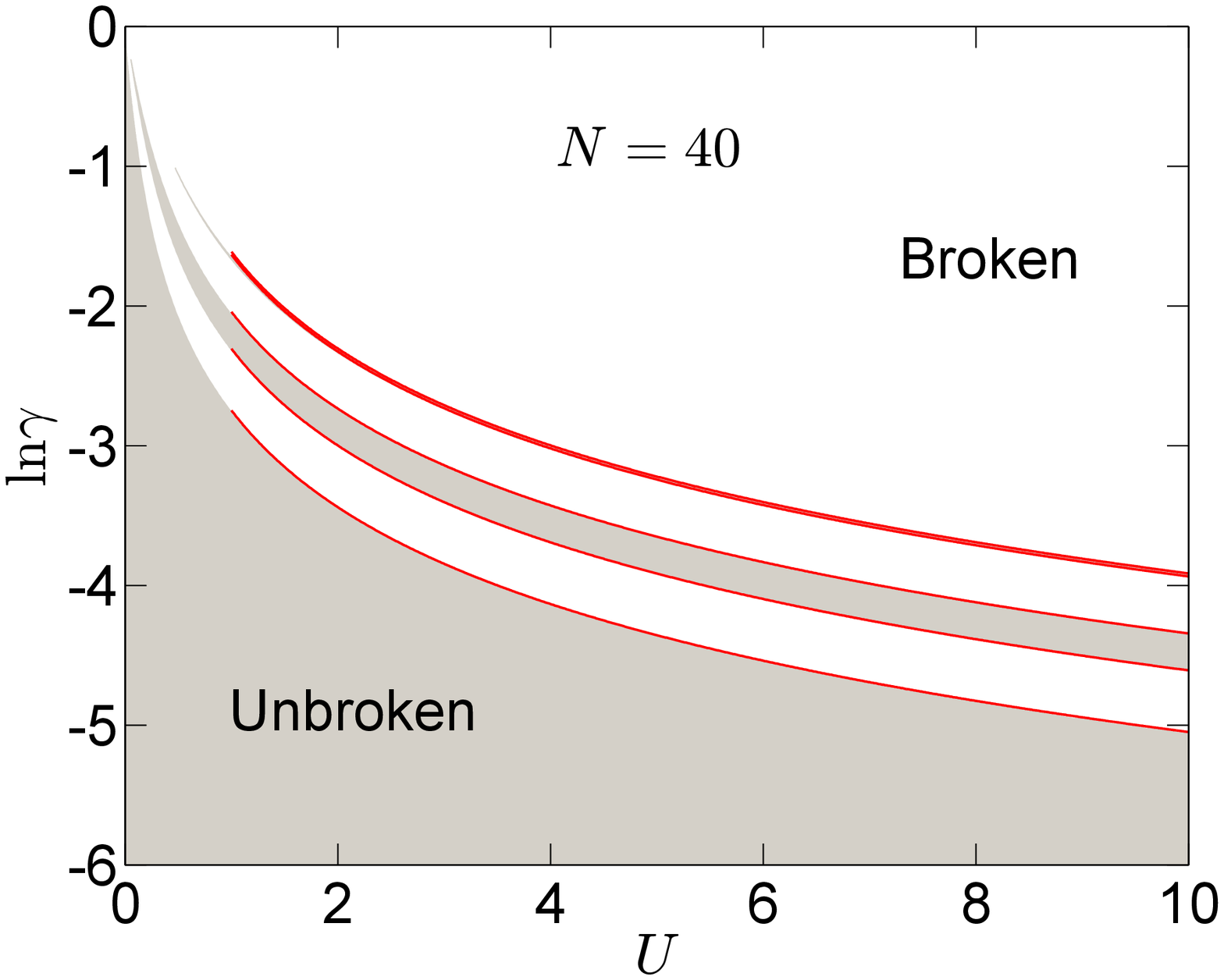}
\caption{The analytical boundaries (\protect\ref{gamma_c_N10}), (\protect\ref%
{gamma_c_N20}), (\protect\ref{gamma_c_N40}) (red lines) and the phase
diagram in large $U$ region obtained by exact diagonalization for system $H$
of size $N=10,20,40$. The analytical boundaries are plotted in the region of
$U=1$ to $10$, where $J$ is set as the unit. It is noticed that the
analytical boundaries fit the exact phase diagram well at the large $U$
region. }
\label{phase_diagramN40}
\end{figure}

Similarly as in Section \ref{sec_nonzero_interaction}, the Hamiltonian $H_{%
\mathrm{eff}}$ in single-particle invariant subspace can be solved by Bethe
ansatz method. We are interested in the bound-pair band, which has the
spectrum $E\left( K\right) \approx U+(4J^{2}/U)\left( 1+\cos K\right) $.
Here, quasimomenta $K$\ of the bound-pair state\ satisfies the equation%
\begin{equation}
g\left( K\right) =2(J^{2}/U)^{2}\sin \left( NK\right) \left( 1+\cos K\right)
+\gamma ^{2}\sin \left[ \left( N-1\right) K\right] =0.\text{ }
\label{CriticalEQ_Heff}
\end{equation}%
As pointed above, together with the condition $\mathrm{d}g\left( K\right) /%
\mathrm{d}K=0$, one can obtain the equation%
\begin{equation}
Z\left( K\right) =N\left( \cos K+1\right) \sin K-\sin \left( NK\right)
\left\{ \cos \left( NK\right) +\cos \left[ \left( N-1\right) K\right]
\right\} =0,\text{ }  \label{CriticalEQ_Heff_reduced}
\end{equation}%
which solutions $K_{c}\in \left( 0,\pi \right) $ determine the boundaries of
quantum phase for the effective Hamiltonian $H_{\mathrm{eff}}$.\ In other
words, function $Z\left( K\right) $ with $\mathcal{N}_{c}$\ zeros indicates $%
\lfloor \mathcal{N}_{c}/2\rfloor +1\ $unbroken $\mathcal{PT}$-symmetric
regions of\ both $H_{\mathrm{eff}}$ and $H$, where $\lfloor \mathcal{N}%
_{c}/2\rfloor $\ denotes the integer part of $\mathcal{N}_{c}/2$. To
demonstrate this point, function $Z\left( K\right) $ is plotted for
different $N$ in Fig. \ref{fig_Zk}. One can see that, the number of
solutions $\mathcal{N}_{c}$ increases as $N$ increases, which corresponds to
the increasing unbroken $\mathcal{PT}$-symmetric regions. To be specific,
for $N=10$, $20$, and $40$, it shows that $\mathcal{N}_{c}=1$, $3$, and $5$,
respectively. This indicates there are $1$, $2$, and $3$\ unbroken regions,
which accords to the phase diagram in Fig. \ref{phase_diagramN40}.
Quantitatively, one can obtain the solutions of $K_{c}$ and $\gamma _{c}$
from (\ref{CriticalEQ_Heff}) for these three cases numerically, which are
listed in the following,%
\begin{eqnarray}
&&%
\begin{tabular}{rr}
$K_{c}/\pi =$ & $0.917579,$ \\
$-\ln \gamma _{c}/\ln (J^{2}/U)=$ & $1.515811,$%
\end{tabular}%
\text{ }(\mathrm{for}\text{ }N=10),  \label{gamma_c_N10} \\
&&%
\begin{tabular}{rrrr}
$K_{c}/\pi =$ & $0.956567,$ & $0.938854,$ & $0.914870,$ \\
$-\ln \gamma _{c}/\ln (J^{2}/U)=$ & $2.124320,$ & $1.510840,$ & $1.395938,$%
\end{tabular}%
\text{ }(\mathrm{for}\text{ }N=20),  \label{gamma_c_N20} \\
&&%
\begin{tabular}{rrrrrr}
$K_{c}/\pi =$ & $0.977429,$ & $0.971547,$ & $0.955142,$ & $0.942098,$ & $%
0.933982,$ \\
$-\ln \gamma _{c}/\ln (J^{2}/U)=$ & $2.746955,$ & $2.305626,$ & $2.041551,$
& $1.633456,$ & $1.611533.$%
\end{tabular}%
\text{ }(\mathrm{for}\text{ }N=40).  \label{gamma_c_N40}
\end{eqnarray}%
According to the above analysis, equations (\ref{gamma_c_N10}, \ref%
{gamma_c_N20}, \ref{gamma_c_N40}) represent the phase boundaries, which are
plotted as red lines in Fig. \ref{phase_diagramN40} as comparison. It shows
that the boundaries obtained from equations (\ref{gamma_c_N10}, \ref%
{gamma_c_N20}, \ref{gamma_c_N40}) are in agreement with the exact phase
diagram well in large $U$ regime.

For large $N$, the number of unbroken regions can be estimated as the
integer around $\sqrt{2N}/\pi $, the analytical results for different $N$\
of Fig. \ref{fig_Zk} is listed in (\ref{region_number}) as comparison, it is
noticed that the analysis accords with the plots in Fig. \ref{fig_Zk}.%
\begin{equation}
\begin{array}{ccccccc}
N & 10 & 20 & 40 & 200 & 1000 & 5000 \\
\sqrt{2N}/\pi & 1.42 & 2.01 & 2.85 & 6.37 & 14.24 & 31.83 \\
\lbrack \sqrt{2N}/\pi ] & 1 & 2 & 3 & 6 & 14 & 32%
\end{array}%
.  \label{region_number}
\end{equation}%
The ceiling (highest)\ phase boundary is determined by the root of (\ref%
{CriticalEQ_Heff_reduced})\ near $K_{0}\approx \sqrt{2/N}$, correspondingly,
the ceiling phase boundary $\gamma _{c}$ is near $\gamma _{0}\approx
(J^{2}/U)\sqrt{2/\left( N-1\right) }$.

\begin{figure}[tbh]
\includegraphics[ bb=20 175 560 600, width=5.5 cm, clip]{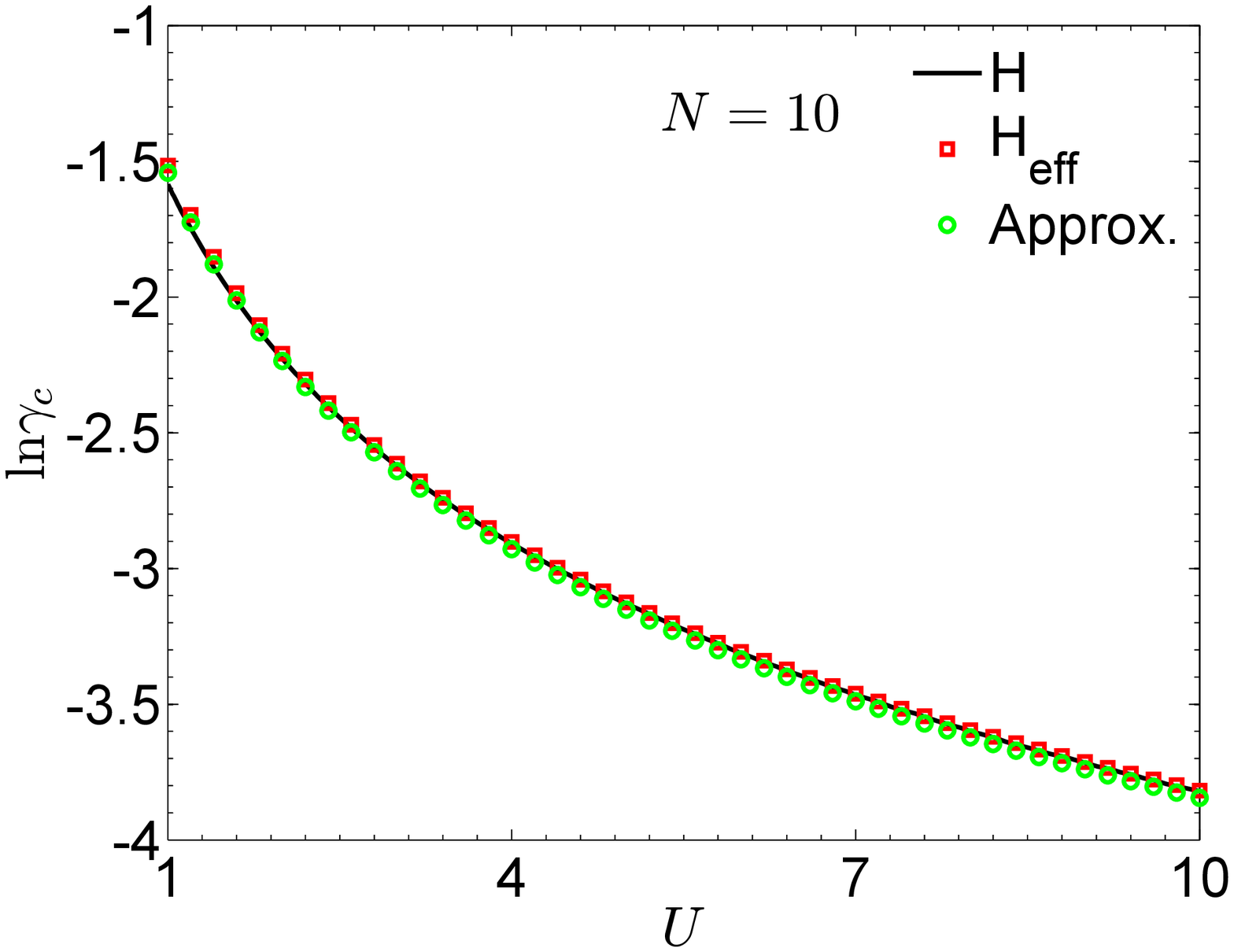} %
\includegraphics[ bb=20 175 560 600, width=5.5 cm, clip]{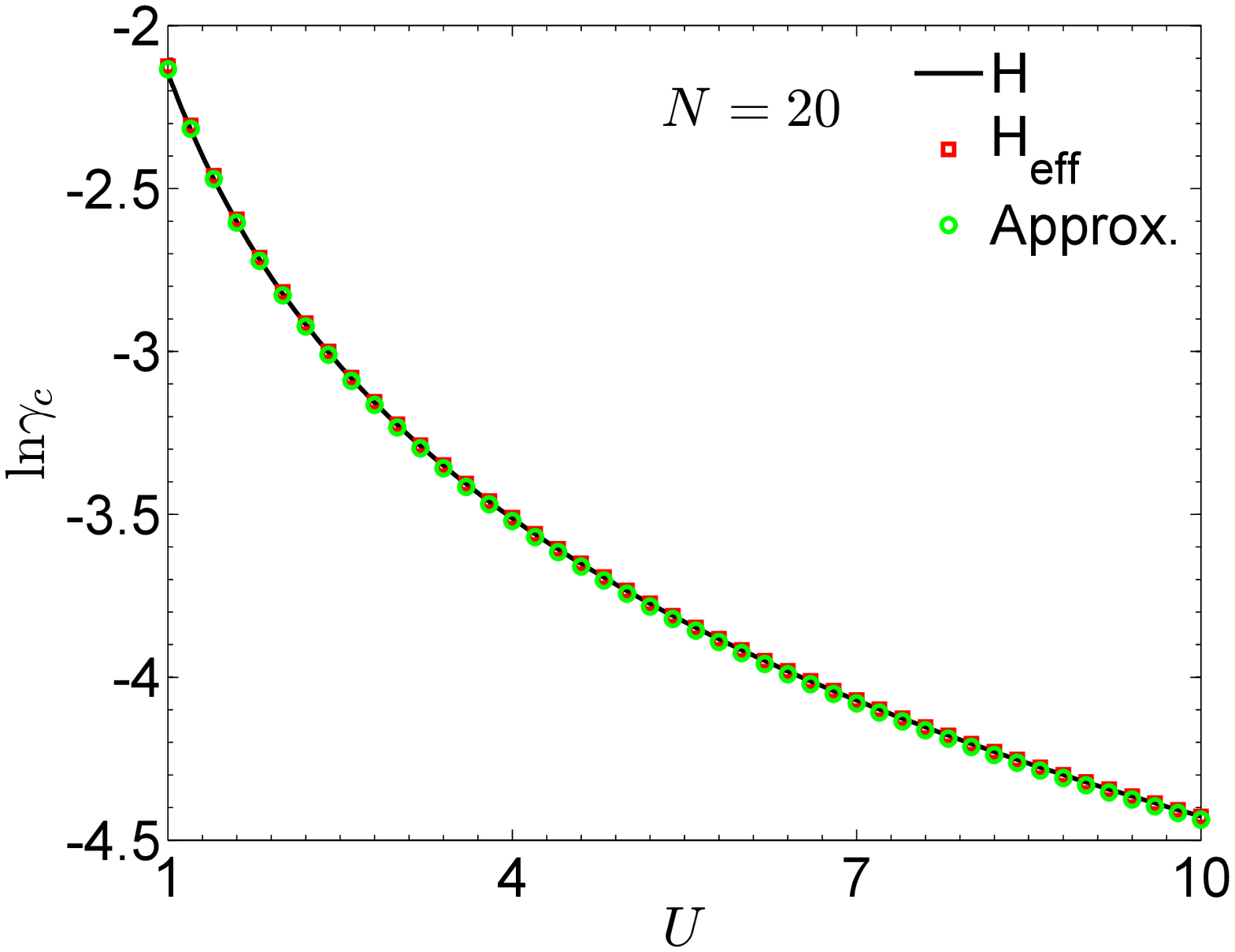} %
\includegraphics[ bb=20 175 560 600, width=5.5 cm, clip]{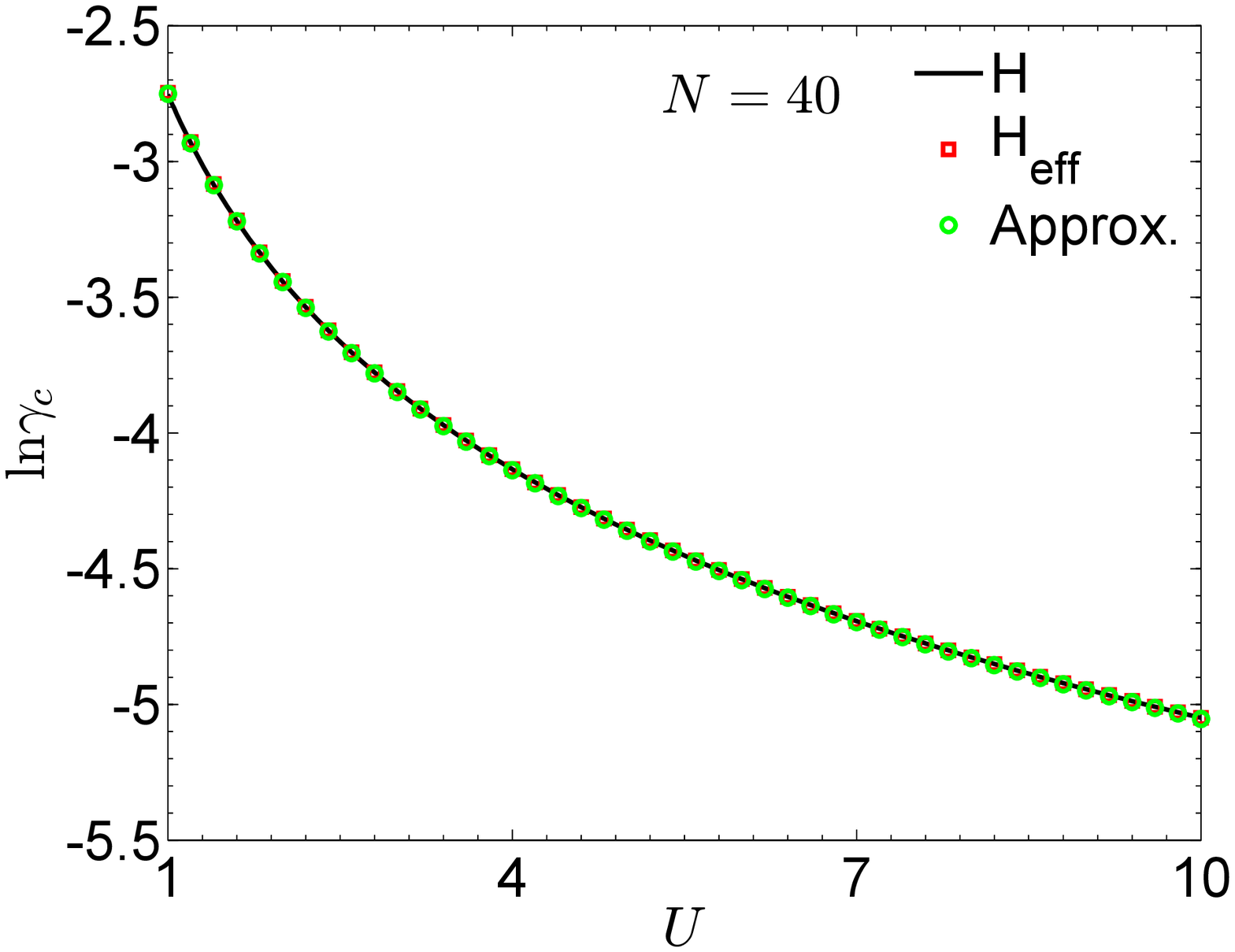}
\caption{ Plots of the floor phase boundary of the phase diagram for $%
N=10,20,40$. Black solid lines are the exact results obtained from
diagonalization of the Bose-Hubbard Hamiltonian $H$. Red squares represent
the exact result obtained from the effective Hamiltonian $H_{\mathrm{eff}}$.
Green circles are the analytical approximation (\protect\ref{gamma_appro}).
It is noticed that $H_{\mathrm{eff}}$ and the approximation effectively
describe the floor phase boundary of $H$.}
\label{fig_comparison}
\end{figure}

Now we turn to analyse the floor (lowest) phase boundary for $H_{\mathrm{eff}%
}$, which corresponds to the solutions $K_{c}$ being in the region $%
[(N-1)\pi /N,\pi ]$. Setting $K=\pi -\left( 1-\delta \right) \pi /N$ with $%
\delta \ll 1$ for large $N$, equation (\ref{CriticalEQ_Heff_reduced}) can be
reduced to%
\begin{equation}
N\left\{ 1-\cos \left[ \left( 1-\delta \right) \pi /N\right] \right\} \sin %
\left[ \left( 1-\delta \right) \pi /N\right] -\sin \left( \pi \delta \right)
\left\{ \cos \left( \pi \delta \right) -\cos \left[ \pi \delta +\left(
1-\delta \right) \pi /N\right] \right\} =0.\text{ }
\end{equation}%
Moreover, by applying the Taylor expansion, it becomes
\begin{equation}
2\left( N-1\right) \delta ^{2}+3\delta -1=0.
\end{equation}%
Solving this equation, one can obtain $\delta _{c}=(\sqrt{8N+1}-3)/\left(
4N-4\right) $, and then the exceptional point $\gamma _{c}$\ as
\begin{equation}
\gamma _{c}\approx \frac{J^{2}}{U}\sqrt{\frac{\pi ^{2}}{N}\frac{\delta
_{c}\left( \delta _{c}-1\right) ^{2}}{\left( N-1\right) \delta _{c}+1}}.
\label{gamma_appro}
\end{equation}%
We plot the exact and analytical approximation results for the floor phase
boundary of $H_{\mathrm{eff}}$, the exact result for the original
Hamiltonian $H$\ in Fig. \ref{fig_comparison} as a comparison. The plots fit
well, especially at large $U$ case. Remarkably, it also gives a good
approximation for those of medium $U$.

\begin{figure}[tbp]
\includegraphics[ bb=30 180 555 590, width=5.6 cm, clip]{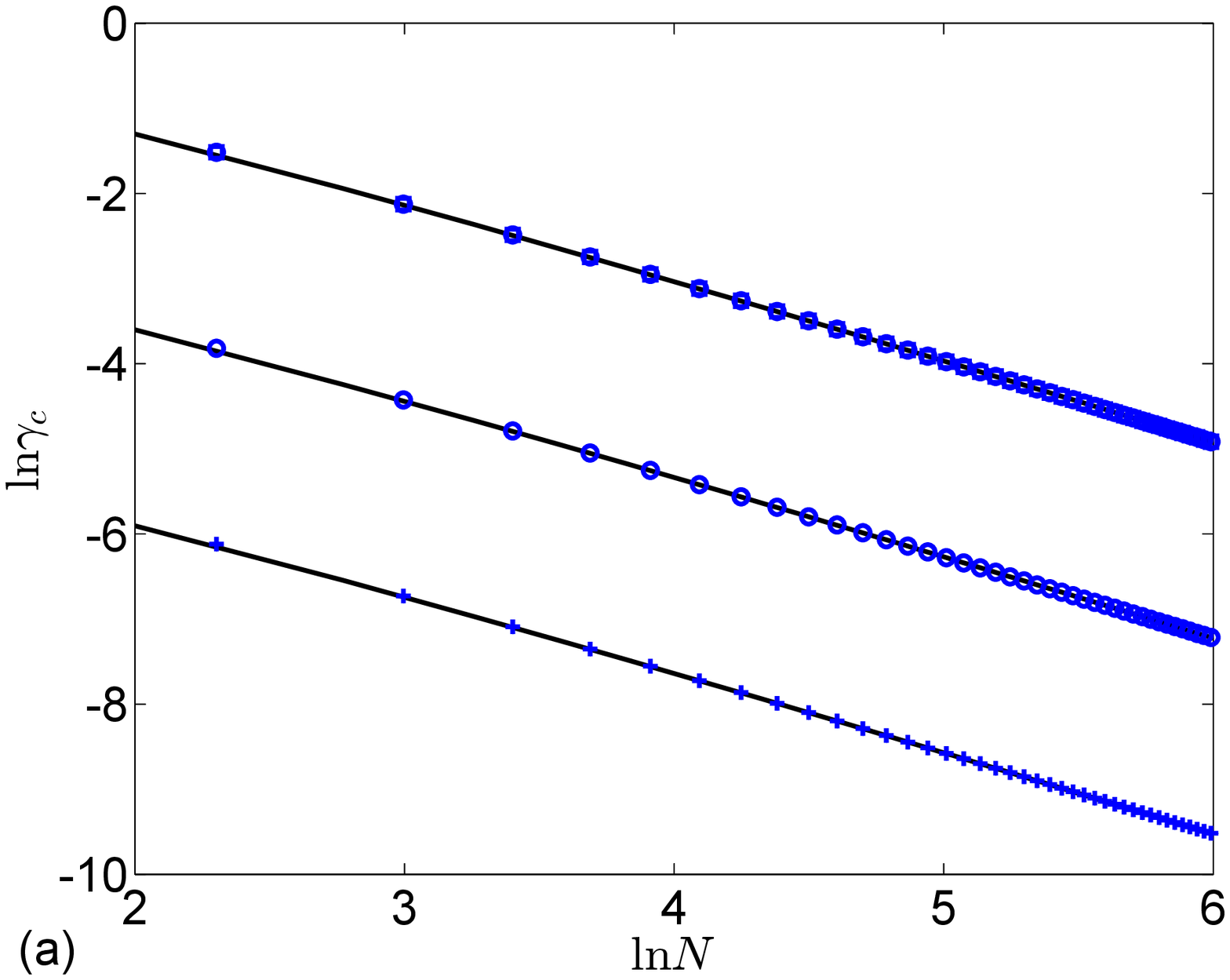} %
\includegraphics[ bb=30 180 555 590, width=5.6 cm, clip]{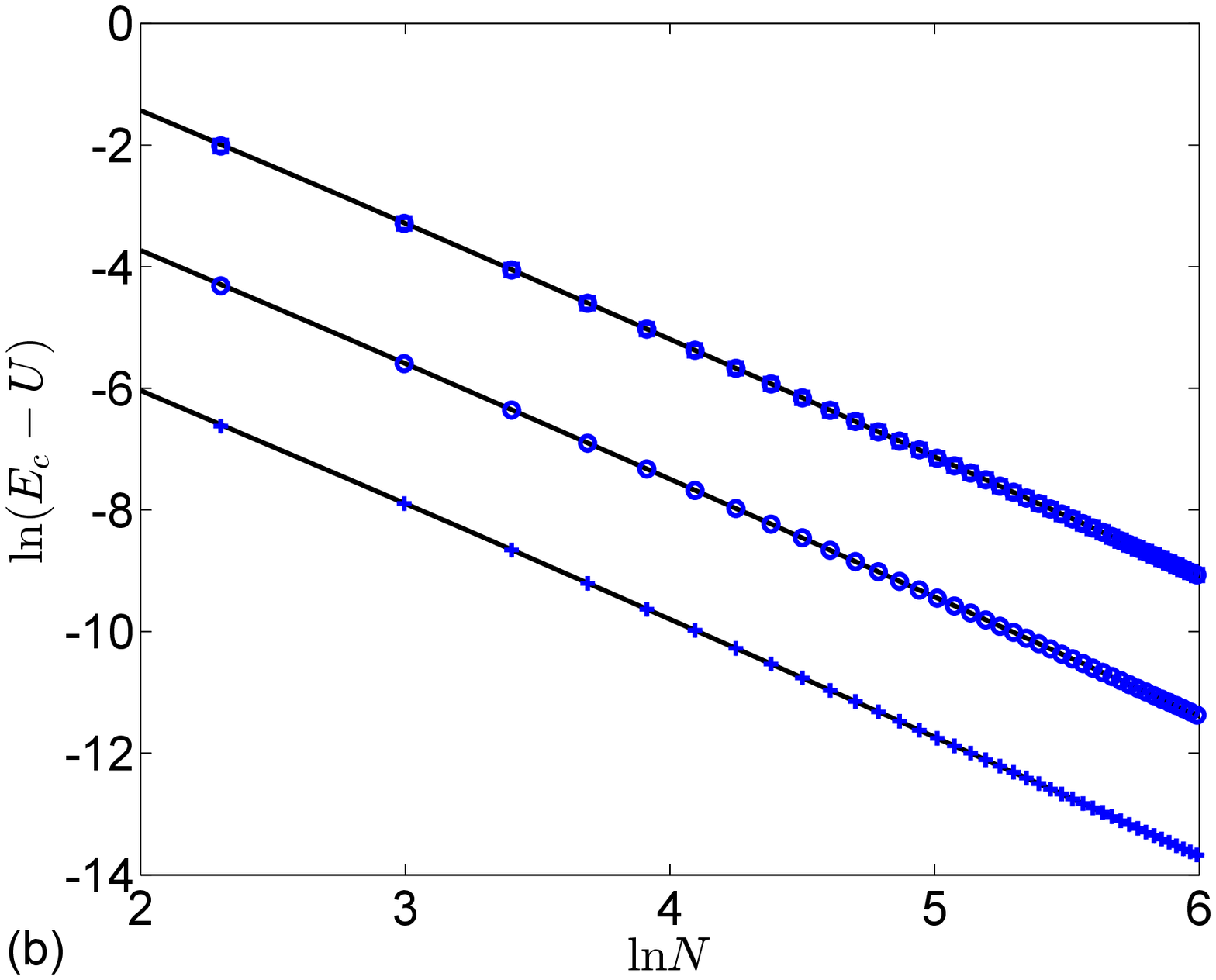} %
\includegraphics[ bb=30 180 555 590, width=5.6 cm, clip]{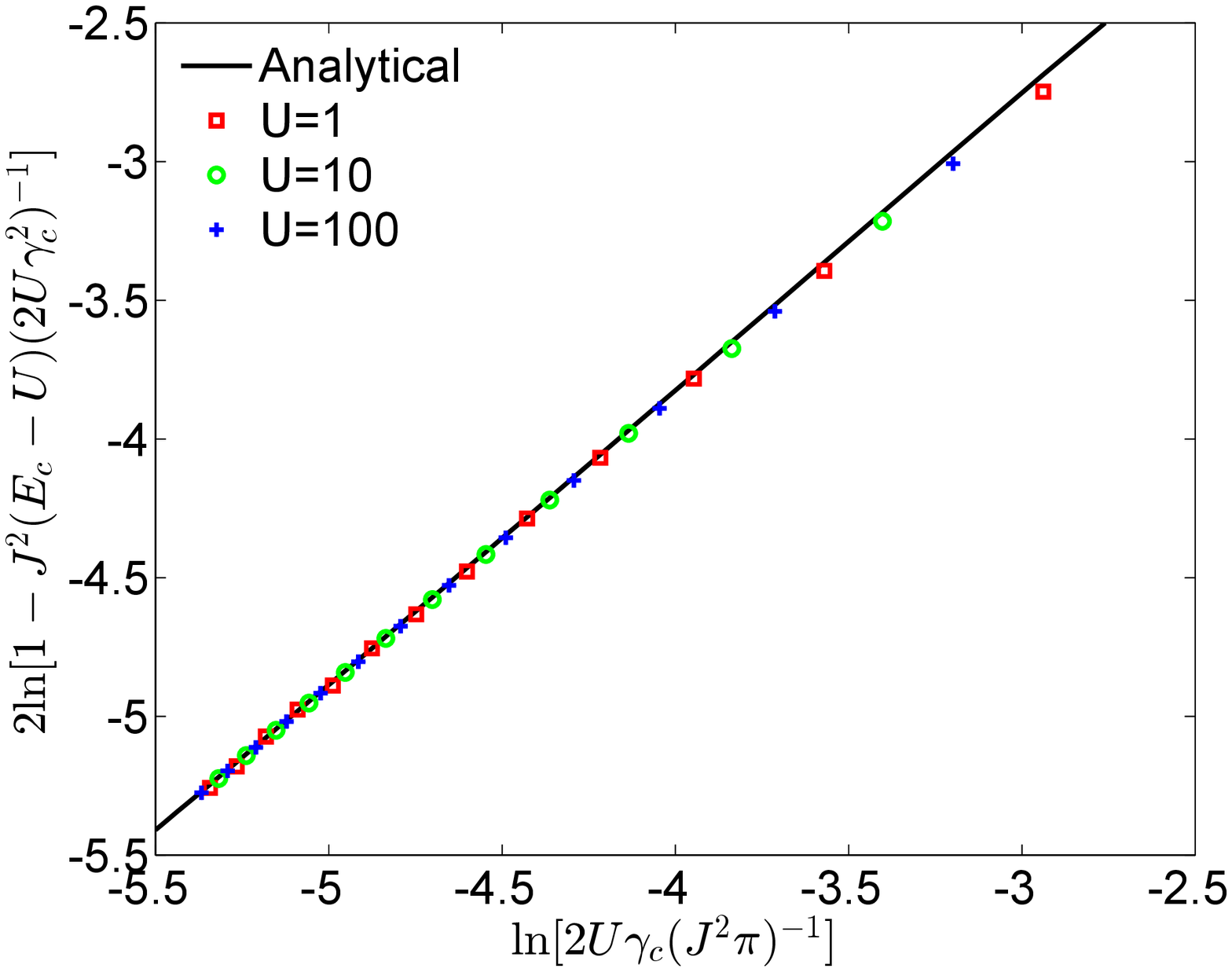}
\caption{Plots of (\protect\ref{gamma_appro}), (\protect\ref{E_c_appro}), (%
\protect\ref{Scaling_appro}) and the corresponding numerical simulation
obtained by exact diagonalization of $H_{\mathrm{eff}}$. $J=1$ is the unit.
In (a), (b), the blue squares, circles and crosses indicate the numerical
results for the cases of $U=1$, $10$ and $100$, respectively. In (c), the
red squares, green circles and blue crosses stand for $U=1$, $10$ and $100$.
The black lines are the plots of the corresponding analytical expressions (%
\protect\ref{gamma_appro}), (\protect\ref{E_c_appro}), (\protect\ref%
{Scaling_appro}). It shows that they are in agreement in large $N$ region.}
\label{fig_scaling_LU}
\end{figure}

It is observed from Fig. \ref{fig_comparison} that $H_{\mathrm{eff}}$ gives
quite good description of the phases of $H$\ for system with $U>1$. On the
other hand, the critical energy $E_{c}$ can be approximately expressed as%
\begin{equation}
E_{c}\approx U+\frac{2J^{2}\pi ^{2}}{UN^{2}}\left( 1-\delta _{c}\right) ^{2}
\label{E_c_appro}
\end{equation}%
for large $N$. From the expression (\ref{gamma_appro}), (\ref{E_c_appro}) of
$\gamma _{c}$, $E_{c}$, we have
\begin{equation}
\left( 1-\frac{J^{2}}{2U}\frac{E_{c}-U}{\gamma _{c}^{2}}\right) ^{2}\approx
\frac{2U\gamma _{c}}{J^{2}\pi },  \label{Scaling_appro}
\end{equation}%
which is a universal scaling law for such a phase transition in large $N$, $%
U $ case. Numerical simulations are performed to investigate the scaling
behavior. We compute the quantities $\gamma _{c}$ and $E_{c}$ for finite
systems $H_{\mathrm{eff}}$, which are plotted in Fig. \ref{fig_scaling_LU}
as comparison with the analytical results (\ref{gamma_appro}), (\ref%
{E_c_appro}) and (\ref{Scaling_appro}). It shows for large $N$, they are in
agreement with each other.%
%

In the limit $N$\ goes to infinity, we notice that $\gamma _{c}=1$\ when $%
U=0 $, means that in the region $\gamma <1$, system $H$ is in the unbroken $%
\mathcal{PT}$-symmetric phase.\ In non-zero interaction ($U\neq 0$) case,
from the above discussions we have the floor phase boundary $\gamma
_{c}\approx J^{2}\pi \left( UN\right) ^{-1}$ and ceiling phase boundary $%
\gamma _{0}\approx (J^{2}/U)\sqrt{2/(N-1)}$ for large $N$.\ As $N$ goes to
infinity, it is noticed that $\lim_{N\rightarrow \infty }\left( \gamma
_{c}\right) =0$ and $\lim_{N\rightarrow \infty }\left( \gamma _{0}\right) =0$
for non-zero finite $U$. In other words, there always exist bound-pair bound
states with conjugate complex energies, which result in the $\mathcal{PT}$%
-symmetry breaking for non-zero $\gamma $.

System features experience huge changes as system parameter approaching the $%
\mathcal{PT}$-symmetric phase transition point (exceptional point).
Distinguished from phase transition in Hermitian system, the non-analytic
properties in $\mathcal{PT}$-symmetric phase transition are caused by the
Hamiltonian becoming a Jordan block operator at the exceptional point.
Correspondingly, dynamical behavior near the exceptional point experiences
dramatic changes, e.g. the power oscillation amplitude becomes significantly
large and the corresponding oscillation frequency becomes small. These
features could be useful for weak signal detection, or for amplifier design.
Phase diagram and the scaling behavior show us rich information where phase
transition happens, this maybe helpful for the design and application of
quantum devices.

\section{Conclusion and discussion}

\label{sec_conclusion} In this paper, we have studied the scaling behavior
and phase diagram of a non-Hermitian $\mathcal{PT}$-symmetric Bose-Hubbard
model. For interaction-free case, the metric operator is constructed, which
is employed to investigate the particle number operator and the
corresponding Hermitian counterpart. The derived properties of the metric
operator, similarity matrix and equivalent Hamiltonian reflect the fact that
they have a common feature: all the matrix elements change dramatically with
diverging derivatives near the exceptional point. For nonzero $U$ case, it
is found that even small on-site interaction can break the $\mathcal{PT}$
symmetry drastically. It has been demonstrated that the scaling behavior can
be established for the exceptional point in both small and large $U$ limit.
Based on numerical approach, we find that the phase diagram shows rich
structure for medium $U$ and analyse the phase transition boundary. Finally,
it is worthwhile to point out that the phase transition discussed differs
from the original quantum phase transition in a Hermitian system~\cite%
{Sachdev}. The former aims at certain eigenstates of a non-Hermitian
Hamiltonian, while the later concerns only the ground state of a Hermitian
one. However, our finding reveals that both of them exhibit the scaling
behavior, which may be due to they both relate to the spontaneous symmetry
breaking.

Finally, we would like to discuss the relevance of the present model to a
real physical system. The Hermitian Bose-Hubbard model is the simplest model
capturing the main physics of not only cold atoms in optical lattice also
photons in nonlinear waveguide \cite{Jaksch,Hartmann,QBEC,LonghiJPB44}. The
effective non-Hermitian Bose Hubbard\ Hamiltonian $H_{\mathrm{Loss}}$,\ is
introduced when the closed system couples to the continuum \cite%
{Mahaux,Rotter,Verbaarschot,Dittes}. A experimental realization of such an
open system could be achieved by tunneling escape of atoms from a
magneto-optical trap \cite{Kepesidis} or using lossy cavities \cite{Scala}.
The Hamiltonian $H_{\mathrm{Loss}}$\ has been investigated by solving the
master equation \cite{Kepesidis,Trimborn,Witthaut} or Schr\"{o}dinger
equation \cite{Graefe08PRL,GraefePRA82}. Although the present model
Hamiltonian $H$ (\ref{H_U}) contains an extra gain term $i\gamma $, it has
connection to a Hamiltonian $H_{\mathrm{Loss}}$ by applying a constant
energy shift:%
\begin{equation}
H_{\mathrm{Loss}}=H-i\gamma \sum_{l=1}^{N}n_{l}.
\end{equation}%
The dynamics of $H_{\mathrm{Loss}}$ can be obtained by solving the following
master equation
\begin{equation}
\dot{\rho}=-i\left[ H_{0},\rho \right] -\gamma
\sum_{l=2}^{N-1}(a_{i}^{\dagger }a_{i}\rho +\rho a_{i}^{\dagger
}a_{i}-2a_{i}\rho a_{i}^{\dagger })-2\gamma (a_{N}^{\dagger }a_{N}\rho +\rho
a_{N}^{\dagger }a_{N}-2a_{N}\rho a_{N}^{\dagger }),  \label{master_eq}
\end{equation}%
where $\rho $\ is the density matrix of $H_{\mathrm{Loss}}$, $%
H_{0}=-J\sum_{l=1}^{N-1}(a_{l}^{\dag }a_{l+1}+$\textrm{H.c.}$%
)+(U/2)\sum_{l=1}^{N}a_{l}^{\dag 2}a_{l}^{2}$.

Based on the result of this paper, some important features of $H_{\mathrm{%
Loss}}$ can be observed. Firstly, although Hamiltonian $H_{\mathrm{Loss}}$\
is not invariant under $\mathcal{PT}$ transformation, the eigenstates of $H_{%
\mathrm{Loss}}$\ are still $\mathcal{PT}$ symmetric within the unbroken
region of $H$. Secondly, two Hamiltonians $H$\ and $H_{\mathrm{Loss}}$\
share the same dynamics except the extra decaying factor, i.e., $\psi \left(
t\right) \rightarrow \psi \left( t\right) e^{-\gamma t}$. From this
perspective, the phase boundary as well as the scaling law presented in this
paper, can be observed from the dynamics of $H_{\mathrm{Loss}}$. As a future
work, it is interesting to compare the results obtained by two methods.
Actually, it has been explored for a two-site example \cite%
{Trimborn,Witthaut}.

\section*{Acknowledgment}

We acknowledge the support of National Basic Research Program (973 Program)
of China under Grant No. 2012CB921900.

\appendix

\section{Operators $\hat{n}_{l}$ and $\hat{n}_{k_{l}}$}

\label{sec_appendix} This appendix provides the examples to demonstrate that
operators $\hat{n}_{l}=a_{l}^{\dag }a_{l}$ and $\hat{n}_{k_{l}}=a_{k_{l}}^{%
\dagger }a_{k_{l}}$ are not observables. We consider the single-particle
case as an illustrative example, the metric operator can be expressed as%
\begin{equation}
\eta =\sum_{k,\mu ,\nu }g_{k}^{\mu }\left( g_{k}^{\nu }\right) ^{\ast
}\left\vert \mu \right\rangle \left\langle \nu \right\vert ,
\end{equation}%
and the operators are $\hat{n}_{l}=\left\vert l\right\rangle \left\langle
l\right\vert $ and $\hat{n}_{k_{l}}=\left\vert k_{l}\right\rangle
\left\langle k_{l}\right\vert $. Then we have
\begin{eqnarray}
(\eta \hat{n}_{l}-\hat{n}_{l}^{\dagger }\eta )_{j_{1}j_{2}} &=&\left\langle
j_{1}\right\vert (\eta \hat{n}_{l}-\hat{n}_{l}^{\dagger }\eta )\left\vert
j_{2}\right\rangle \\
&=&\sum_{k}\left[ g_{k}^{j_{1}}(g_{k}^{l})^{\ast }\delta
_{l,j_{2}}-g_{k}^{l}(g_{k}^{j_{2}})^{\ast }\delta _{l,j_{1}}\right]  \notag
\end{eqnarray}%
or more explicitly
\begin{equation}
(\eta \hat{n}_{l}-\hat{n}_{l}^{\dagger }\eta )_{j_{1}j_{2}}=\left\{
\begin{array}{cc}
-\sum_{k}g_{k}^{l}(g_{k}^{j_{2}})^{\ast }, & \left( j_{1}=l\neq j_{2}\right)
\\
\sum_{k}\left( g_{k}^{l}\right) ^{\ast }g_{k}^{j_{1}}, & \left( j_{1}\neq
l=j_{2}\right) \\
0, &
\begin{array}{c}
j_{1}=l=j_{2} \\
\text{or }j_{1}\neq l\neq j_{2}%
\end{array}%
\end{array}%
\right. .
\end{equation}%
We note that $\sum_{k}(g_{k}^{j_{1}})^{\ast }g_{k}^{j_{2}}$ do not vanish in
general case, which can be seen in the following illustrative example. We
consider the case with $N=2$, from the operator $\eta _{2}$ listed in the
Table \ref{table}, we notice that%
\begin{equation}
\begin{array}{c}
\eta _{2}\hat{n}_{1}-\hat{n}_{1}^{\dagger }\eta _{2}=\frac{-\gamma }{\sqrt{%
J^{2}-\gamma ^{2}}}\left(
\begin{array}{cc}
0 & i \\
i & 0%
\end{array}%
\right) \neq 0, \\
\eta _{2}\hat{n}_{2}-\hat{n}_{2}^{\dagger }\eta _{2}=\frac{\gamma }{\sqrt{%
J^{2}-\gamma ^{2}}}\left(
\begin{array}{cc}
0 & i \\
i & 0%
\end{array}%
\right) \neq 0,%
\end{array}%
\end{equation}%
which means $\eta _{2}\hat{n}_{l}\eta _{2}^{-1}\neq \hat{n}_{l}^{\dagger }$.
Accordingly, it leads to
\begin{equation}
\eta (l\hat{n}_{l})\eta ^{-1}=l\eta \hat{n}_{l}\eta ^{-1}\neq l\hat{n}%
_{l}\eta \eta ^{-1}=(l\hat{n}_{l})^{\dagger },
\end{equation}%
which shows that the position operator $l\hat{n}_{l}$ is not an observable.
This accords with the conclusion of Ref.~\cite{AM37,BenderRPP,Heiss40}.

Similarly, the operator $\hat{n}_{k_{l}}$, for $N=2$ system in
single-particle case, has the form%
\begin{equation}
\begin{array}{c}
\hat{n}_{k_{1}}=\frac{-1}{2\sqrt{J^{2}-\gamma ^{2}}}\left(
\begin{array}{cc}
-J & \sqrt{J^{2}-\gamma ^{2}}-i\gamma \\
\sqrt{J^{2}-\gamma ^{2}}+i\gamma & -J%
\end{array}%
\right) , \\
\hat{n}_{k_{2}}=\frac{1}{2\sqrt{J^{2}-\gamma ^{2}}}\left(
\begin{array}{cc}
J & \sqrt{J^{2}-\gamma ^{2}}+i\gamma \\
\sqrt{J^{2}-\gamma ^{2}}-i\gamma & J%
\end{array}%
\right) .%
\end{array}%
\end{equation}%
in coordinate space. Straightforward algebra shows
\begin{equation}
\begin{array}{c}
\eta _{2}\hat{n}_{k_{1}}-\hat{n}_{k_{1}}^{\dagger }\eta _{2}=\frac{\gamma }{%
\sqrt{J^{2}-\gamma ^{2}}}\left(
\begin{array}{cc}
-i & 0 \\
0 & i%
\end{array}%
\right) \neq 0, \\
\eta _{2}\hat{n}_{k_{2}}-\hat{n}_{k_{2}}^{\dagger }\eta _{2}=\frac{\gamma }{%
\sqrt{J^{2}-\gamma ^{2}}}\left(
\begin{array}{cc}
i & 0 \\
0 & -i%
\end{array}%
\right) \neq 0,%
\end{array}%
\end{equation}%
which means $\eta _{2}\hat{n}_{k_{l}}\eta _{2}^{-1}\neq \hat{n}%
_{k_{l}}^{\dagger }$. Then we conclude that operator $\hat{n}_{k_{l}}$ is
not an observable.


\begin{thebibliography}{99}
\bibitem{nonHermitianH} H. Feshbach, Ann. Phys. (N. Y.) 5 (1958) 357;

\nonumber H. Feshbach, Ann. Phys. (N. Y.) 19 (1962) 287;

\nonumber J. Oko\l owicz, M. P\l oszajczaka, I. Rotter, Phys. Rep. 374
(2003) 271;

\nonumber J.G. Muga, J.P. Palao, B. Navarro, I.L. Egusquiza, Phys. Rep. 395
(2004) 357;

\nonumber E.J. Br\"{a}ndas, E.S. Kryachko (Eds.), Fundamental World of
Quantum Chemistry, Vol. II, Kluwer Academic Publishers, Dordrecht, The
Netherlands, 2003.

\bibitem{Geyer1992} F.G. Scholtz, H.B. Geyer, F.J.W. Hahne, Ann. Phys. (NY)
213 (1992) 74.

\bibitem{AliM} A. Mostafazadeh, J. Math. Phys. 43 (2002) 205;

\nonumber A. Mostafazadeh, Int. J. Geom. Meth. Mod. Phys. 7 (2010) 1191.

\bibitem{AM37} A. Mostafazadeh, A. Batal, J. Phys. A: Math. Gen. 37 (2004)
11645.

\bibitem{AM38} A. Mostafazadeh, J. Phys. A: Math. Gen. 38 (2005) 6557.

\bibitem{Ahmed} Z. Ahmed, Phys. Lett. A 282 (2001) 343;

\nonumber Z. Ahmed, Phys. Lett. A 286 (2001) 30;

\nonumber Z. Ahmed, Phys. Rev. A 64 (2001) 042716.

\bibitem{Berry} M.V. Berry, J. Phys. A 31 (1998) 3493;

\nonumber M.V. Berry, Czech. J. Phys. 54 (2004) 1039.

\bibitem{EPsHeiss} W.D. Heiss, A.L. Sannino, J. Phys. A 23 (1990) 1167;

\nonumber W.D. Heiss, Phys. Rep. 242 (1994) 443;

\nonumber W.D. Heiss, J. Phys. A 37 (2004) 2455;

\nonumber F. Leyvraz, W.D. Heiss, Phys. Rev. Lett. 95 (2005) 050402.

\bibitem{Dembowski} C. Dembowski, H.-D. Gr\"{a}f, H.L. Harney, A. Heine,
W.D. Heiss, H. Rehfeld, A. Richter, Phys. Rev. Lett. 86 (2001) 787;

\nonumber C. Dembowski, B. Dietz, H.-D. Gr\"{a}f, H.L. Harney, A. Heine,
W.D. Heiss, A. Richter, Phys. Rev. E 69 (2004) 056216.

\bibitem{Heiss40} D.P. Musumbu, H.B. Geyer, W.D. Heiss, J. Phys. A: Math.
Theor. 40 (2007) F75.

\bibitem{Bender98} C.M. Bender, S. Boettcher, Phys. Rev. Lett. 80 (1998)
5243.

\bibitem{Bender} C.M. Bender, D.C. Brody, H.F. Jones, Phys. Rev. Lett. 89
(2002) 270401;

\nonumber C.M. Bender, D.C. Brody, H.F. Jones, B.K. Meister, Phys. Rev.
Lett. 98 (2007) 040403;

\nonumber C.M. Bender, P.D. Mannheim, Phys. Rev. Lett. 100 (2008) 110402;

\nonumber C.M. Bender, D.W. Hook, P.N. Meisinger, Q.H. Wang, Phys. Rev.
Lett. 104 (2010) 061601.

\bibitem{BenderRPP} C.M. Bender, Rep. Prog. Phys. 70 (2007) 947.

\bibitem{MZnojil} M. Znojil, J. Phys. A: Math. Theor. 40 (2007) 13131;

\nonumber M. Znojil, J. Phys. A: Math. Theor. 41 (2008) 292002;

\nonumber M. Znojil, Phys. Rev. A 82 (2010) 052113

\nonumber M. Znojil, J. Phys. A: Math. Theor. (2011) 075302;

\nonumber F. Bagarello, M. Znojil, J. Phys. A: Math. Theor. 44 (2011) 415305.

\bibitem{Jones} H.F. Jones, J. Phys. A: Math. Gen. 38 (2005) 1741;

\nonumber H.F. Jones, Phys. Rev. D 76 (2007) 125003;

\nonumber H.F. Jones, Phys. Rev. D 78 (2008) 065032.

\bibitem{Tateo} P. Dorey, C. Dunning, R. Tateo, J. Phys. A: Math. Theor. 40
(2007) R205.

\bibitem{Mueller} M. M\"{u}ller, I. Rotter, J. Phys. A: Math. Theor. 41
(2008) 244018.

\bibitem{Longhi} S. Longhi, Phys. Rev. B 80 (2009) 235102;

\nonumber S. Longhi, Phys. Rev. B 81 (2010) 195118;

\nonumber S. Longhi, Phys. Rev. B 82 (2010) 041106(R);

\nonumber S. Longhi, Phys. Rev. A 82 (2010) 032111.

\bibitem{LonghiPRL103} S. Longhi, Phys. Rev. Lett. 103 (2009) 123601.

\bibitem{PTinOptics} R. El-Ganainy, K.G. Makris, D.N. Christodoulides, Z.H.
Musslimani, Opt. Lett. 32 (2007) 2632;

\nonumber Z.H. Musslimani, K.G. Makris, R. El-Ganainy, D.N. Christodoulides,
Phys. Rev. Lett. 100 (2008) 030402;

\nonumber K.G. Makris, R. El-Ganainy, D.N. Christodoulides, Z.H. Musslimani,
Phys. Rev. A 81 (2010) 063807.

\bibitem{PTinOpticsPO} K.G. Makris, R. El-Ganainy, D.N. Christodoulides,
Z.H. Musslimani, Phys. Rev. Lett. 100 (2008) 103904.

\bibitem{Klaiman} S. Klaiman, U. G\"{u}nther, N. Moiseyev, Phys. Rev. Lett.
101 (2008) 080402.

\bibitem{AGuo} A. Guo, G.J. Salamo, D. Duchesne, R. Morandotti, M.
Volatier-Ravat, V. Aimez, G.A. Siviloglou, D.N. Christodoulides, Phys. Rev.
Lett. 103 (2009) 093902;

\nonumber C.E. R\"{u}ter, K.G. Makris, R. El-Ganainy, D.N. Christodoulides,
M. Segev, D. Kip, Nat. Phys. 6 (2010) 192;

\nonumber T. Kottos, Nat. Phys. 6 (2010) 166.

\bibitem{JLPT} L. Jin, Z. Song, Phys. Rev. A 80 (2009) 052107.

\bibitem{Ghosh} C. Korff, R. Weston, J. Phys. A: Math. Theor. 40 (2007) 8845;

\nonumber T. Deguchi, P.K. Ghosh J. Phys. A: Math. Theor. 42 (2009) 475208;

\nonumber O.A. Castro-Alvaredo, A. Fring, J. Phys. A: Math. Theor. 42 (2009)
465211;

\nonumber\"{O}zlem Ye\c{s}ilta\c{s}, J. Phys. A: Math. Theor. 44 (2011)
305305;

\nonumber L.B. Drissi, E.H. Saidi, M. Bousmina, J. Math. Phys. 52 (2011)
022306;

\nonumber L. Jin, Z. Song, Phys. Rev. A 81 (2010) 032109;

\nonumber L. Jin, Z. Song, Phys. Rev. A 83, (2011) 062118;

\nonumber L. Jin, Z. Song, Phys. Rev. A 84 (2011) 042116;

\nonumber L. Jin, Z. Song, J. Phys. A 44 (2011) 375304.

\bibitem{Giorgi} G.L. Giorgi, Phys. Rev. B 82 (2010) 052404.

\bibitem{Bendix} O. Bendix, R. Fleischmann, T. Kottos, B. Shapiro, Phys.
Rev. Lett. 103 (2009) 030402.

\bibitem{Joglekar} Y.N. Joglekar, D. Scott, M. Babbey, A. Saxena Phys. Rev.
A 82 (2010) 030103(R);

\nonumber Y.N. Joglekar, A. Saxena, Phys. Rev. A 83 (2011) 050101(R).

\bibitem{JoglekarDegree} D.D. Scott, Y.N. Joglekar, Phys. Rev. A 83 (2011)
050102(R).

\bibitem{Greiner} M. Greiner, O. Mandel, T. Esslinger, T.W. H\"{a}nsch, I.
Bloch, Nature (London) 415 (2002) 39.

\bibitem{Jaksch} D. Jaksch, C. Bruder, J.I. Cirac, C.W. Gardiner, P. Zoller,
Phys. Rev. Lett. 81 (1998) 3108.

\bibitem{Hartmann} M.J. Hartmann, F.G.S.L. Brand\~{a}o, M.B. Plenio, Nat.
Phys. 2 (2006) 849;

\nonumber M.J. Hartmann, M.B. Plenio, Phys. Rev. Lett. 99 (2007) 103601.

\bibitem{QBEC} M. Greiner, O. Mandel, T. Esslinger, T.W. H\"{a}sch, I.
Bloch, Nature 415 (2002) 39.

\bibitem{LonghiJPB44} S. Longhi, J. Phys. B: At. Mol. Opt. Phys. 44 (2011)
051001.

\bibitem{Hiller} M. Hiller, T. Kottos, A. Ossipov, Phys. Rev. A 73 (2006)
063625.

\bibitem{GraefeJPA41} E.M. Graefe, U. G\"{u}nther, H.J. Korsch, A.E.
Niederle, J. Phys. A 41 (2008) 255206.

\bibitem{Graefe08PRL} E.M. Graefe, H.J. Korsch, A.E. Niederle, Phys. Rev.
Lett. 101 (2008) 150408.

\bibitem{GraefePRA82} E.M. Graefe, H.J. Korsch, A.E. Niederle, Phys. Rev. A
82 (2010) 013629.

\bibitem{HXiong} H. Xiong, Phys. Rev. A 82 (2010) 053615.

\bibitem{HZhong} H. Zhong, W. Hai, G. Lu, Z. Li, Phys. Rev. A 84 (2011)
013410.

\bibitem{Kepesidis} K.V. Kepesidis, M.J. Hartmann, Phys. Rev. A 85 (2012)
063620.

\bibitem{Lindblad} G. Lindblad, Commun. Math. Phys. 48 (1976) 119.

\bibitem{Prosen} T. Prosen, Phys. Rev. Lett. 109 (2012) 090404.

\bibitem{scaling book} M.N. Barber, Phase Transition and Critical Phenomena,
C. Domb, J.L. Lebowitz (Eds.), Academic, New York, Vol. 8, P. 145. 1983.

\bibitem{Winkler} K. Winkler, G. Thalhammer, F. Lang, R. Grimm, J.H.
Denschlag, A.J. Daley, A. Kantian, H.P. B\"{u}chler, P. Zoller, Nature
(London) 441 (2006) 853.

\bibitem{JLBP} L. Jin, B. Chen, Z. Song, Phys. Rev. A 79 (2009) 032108;

\nonumber L. Jin, Z. Song, New J. Phys. 13 (2011) 063009.

\bibitem{JLPerfetBP} L. Jin, Z. Song, Phys. Rev. A 83 (2011) 052102.

\bibitem{BlochRMP80} I. Bloch, J. Dalibard, W. Zwerger, Rev. Mod. Phys. 80
(2008) 885.

\bibitem{Sachdev} S. Sachdev, Quantum Phase Transition, Cambridge University
Press, Cambridge, England, 1999.

\bibitem{Mahaux} C. Mahaux, H.A Weidenm\"{u}ller, Shell Model Approach in
Nuclear Reactions (North-Holland, Amsterdam, 1969).

\bibitem{Rotter} I. Rotter, Rep. Prog. Phys. 54 (1991) 635.

\bibitem{Verbaarschot} J.J.M. Verbaarschot, H.A. Weidenm\"{u}ller, M.R.
Zirnbauer, Phys. Rep. 129 (1985) 367.

\bibitem{Dittes} F. Dittes, Phys. Rep. 339 (2000) 215.

\bibitem{Scala} M. Scala, B. Militello, A. Messina, J. Piilo, S. Maniscalco,
Phys. Rev. A 75 (2007) 013811.

\bibitem{Trimborn} F. Trimborn, D. Witthaut, S. Wimberger, J. Phys. B: At.
Mol. Opt. Phys. 41 (2008) 171001.

\bibitem{Witthaut} D. Witthaut, F. Trimborn, H. Hennig, G. Kordas, T.
Geisel, S. Wimberger, Phys. Rev. A 83 (2011) 063608.
\end{thebibliography}
\end{document}